\documentclass[angeo]{copernicus}
\nolinenumbers

\usepackage{subcaption}
\usepackage{graphicx}
\usepackage{newtxmath}

\begin{document}

\title{Effect of interplanetary shock waves on turbulence parameters}

\Author[1]{Emilia}{Kilpua}
\Author[1]{Simon}{Good}
\Author[1]{Juska}{Soljento}
\Author[2,3]{Domenico}{Trotta}
\Author[1]{Tia}{Bäcker}
\Author[1]{Julia}{Ruohotie}
\Author[1]{Jens}{Pomoell}
\Author[1,4]{Chaitanya}{Sishtla}
\Author[5]{Rami}{Vainio}

\affil[1]{Department of Physics, University of Helsinki, P.O. Box 64, FI-00014 Helsinki, Finland}
\affil[2]{The Blackett Laboratory, Department of Physics, Imperial College London, London SW7 2AZ, UK}
\affil[3]{European Space Astronomy Centre, European Space Agency, Camino Bajo del Castillo s/n, 28692 Villanueva de la Cañada, Madrid, Spain}
\affil[4]{Department of Physics and Astronomy, Queen Mary University of London, London E1 4NS, UK}
\affil[5]{Department of Physics and Astronomy, University of Turku,  FI-20014 Turku, Finland}

\correspondence{Emilia Kilpua (emilia.kilpua@helsinki.fi)}

\runningtitle{Shock effect on turbulence parameters}

\runningauthor{Kilpua et al.}

\firstpage{1}

\maketitle

\begin{abstract}
We have performed an extensive statistical investigation of how interplanetary fast forward shocks affect certain turbulence parameters, namely, the normalised cross-helicity, $\sigma_c$, residual energy, $\sigma_r$, and magnetic helicity, $\sigma_m$.  A total of 371 shocks detected by \textit{Wind} at 1~au and seven shocks by \textit{Solar Orbiter} at 0.3--0.5~au have been analysed. We explore how the aforementioned turbulence parameters and their variation across the shock depend on the shock characteristics, parametrised in terms of the gas compression ratio, upstream plasma beta, velocity jump and shock angle. In the shock vicinity, fluctuations tend on average to show anti-sunward imbalance (measured as positive $\sigma_c$ when rectified to the Parker spiral direction), a dominance of magnetic energy (negative $\sigma_r$) and zero $\sigma_m$, all being typical properties of the solar wind. Anti-sunward imbalance and equipartition ($\sigma_r \sim 0$) in the upstream is increasingly prevalent with increasing shock velocity jump and decreasing upstream beta and shock angle. Shocks with large velocity jumps and gas compression ratios have considerably more balanced ($\sigma_c\sim 0$) and more magnetically dominated fluctuations downstream than upstream. From upstream to downstream, we also find that the occurrence of time periods fulfilling strict criteria for Alfv\'enic fluctuations (AFs) usually decreases, while those meeting the criteria for small-scale flux ropes (SFRs) increases. The occurrence of AF-like periods peaks for quasi-parallel shocks with large velocity jumps and small upstream beta values. The occurrence of SFRs increases with increasing gas compression ratio and upstream beta. The shocks observed by \textit{Solar Orbiter} below 0.5~au display similar distributions of turbulence parameters and upstream-to-downstream changes to those detected at 1~au. These results are relevant for understanding turbulence and charged-particle acceleration at collisionless shocks. 

\end{abstract}

\introduction
\label{sec:introduction}

The solar wind is a continuous flow of collisionless plasma from the Sun into interplanetary space \citep{Parker1958}. It is permeated by a variety of waves, structures and a turbulence cascade in which energy injected at large scales transfers through an intermediate inertial range to kinetic scales, where it finally dissipates \citep[e.g.,][]{Bruno2013,Verscharen2019}. The dominant fluctuation modes in the solar wind are Alfv\'enic in nature \citep{Belcher1971}, exhibiting strong correlation or anti-correlation between the magnetic field and velocity vectors. Alfv\'enic fluctuations are particularly dominant in fast solar wind streams \citep[e.g.,][]{Snekvik2013} but are also observed in the slow solar wind \citep[e.g.,][]{Amicis2021}. The origin of the anti-sunward-propagating Alfv\'en waves seen at injection scales, which supply energy to the turbulent cascade at inertial scales, has been linked, for example, to convective motions in the photosphere, with the waves resulting from these motions being swept into interplanetary space with the solar wind outflow \citep[e.g.,][]{Tomczyk2009,Cranmer2005}. 

Sunward propagating Alfv\'en waves may be generated by reflection of outward propagating waves in the stratified plasma and via non-linear processes such as the parametric decay instability \citep[e.g.,][]{goldstein1978,gary2001,Sishtla2022}, and also in regions where large velocity shears are present \citep[e.g.,][]{Soljento2023}. The sunward fluctuations below and above the critical point must in turn be generated in the sub-Alfvénic and super-Alfvénic solar wind, respectively.

An important question is how fast forward interplanetary shock waves affect the parameters that characterise solar wind fluctuations \citep[e.g.,][]{Zank2021}. Interplanetary shocks are ubiquitous in the solar wind \citep[e.g.][]{Kilpua2015}, and are usually driven by interplanetary coronal mass ejections \citep[ICMEs; e.g.,][]{Kilpua2017a} or they are formed ahead of fast--slow stream interaction regions \citep[e.g.,][]{Richardson2018,Jian2006}. Several statistical analyses have been conducted to investigated how shock waves affect fluctuation power levels and spectral slopes \citep[e.g.,][]{Park2023,Borovsky2020,Pitna2016,Kilpua2021Fr,Pitna2021}, including also theoretical considerations \citep[e.g.,][]{Zank2021}. They have shown that the fluctuation power (and power normalised to the mean background field, to a lesser extent) is enhanced from the upstream to downstream, while spectral slopes both in the inertial and ion dissipation ranges are unaffected or steepen somewhat. The enhancement of fluctuation power is related to the compression at the shock and possible generation of new fluctuations downstream of the shock. The steeper spectral slopes downstream could be related to an increase of intermittent structures such as current sheets. The steepening of the spectral slope (or its invariance) at the shock transition is in contrast to observations at planetary bow shocks, where in some cases an $f^{-1}$ range is found in the downstream, suggesting that the turbulence spectrum is reset \citep[][]{hadid2015,huang2017,huang2020}. Such ``fresh injection'' in the shock downstream has also been modelled by means of kinetic simulations of shocks interacting with laminar and turbulent plasma, showing its relevance at small scales~\citep{Trotta2023b}.

Also among the key parameters for characterising turbulence are the normalised cross-helicity, $\sigma_c$, residual energy, $\sigma_r$, and magnetic helicity, $\sigma_m$ \citep[e.g.,][]{Matthaeus1982,Roberts1987}. The normalised cross-helicity and residual energy may be interpreted as, respectively, the balance in power between Alfv\'enic fluctuations propagating parallel and anti-parallel to the mean magnetic field, and how energy is divided between kinetic (i.e., velocity) and magnetic field fluctuations \citep[][]{Bavassano1998,Verscharen2019,Bruno2013}. These parameters may thus be used to describe `Alfv\'enicity' in the solar wind and they also affect the energisation of charged particles at interplanetary shock waves \citep{vainio1998} by affecting particle scattering close to the shock: if turbulence is Alfv\'enic, normalised cross-helicity will (i) determine the effective scattering-centre speed relative to the medium and, thus, the so-called scattering-centre compression ratio at the shock, which governs first-order Fermi acceleration at the shock, and (ii) the rate of second-order Fermi acceleration, which does not operate in a unidirectional Alfv\'en wave field. The normalised magnetic helicity \citep[][]{Matthaeus1982} is defined as the ratio of the magnetic helicity and total energy spectra. Its value is zero for linearly polarised Alfv\'en wave turbulence at MHD scales (in contrast to the kinetic range). It has been shown, both in the interplanetary case~\citep{Ruohotie2022} and at Earth's bow shock~\citep{Trotta2022b}, that flux ropes transmitted across shocks increase their magnetic helicity content, with important consequences for a series of key phenomena like particle acceleration~\citep{Kilpua2023}.

Near-ecliptic \textit{Parker Solar Probe} observations of the solar wind within 1~au show that the cross-helicity of inertial range fluctuations approaches zero with increasing heliocentric distance from the Sun, while residual energy shows less clear radial trends \citep{Chen2020,Shi2021,Sioulas2023}. The \textit{Ulysses} polar solar wind observations at distances between 1.4 and 4.3~au showed similarly that inertial range fluctuations become more balanced (i.e. have lower cross-helicity) with increasing distance from the Sun, while also becoming less equipartitioned (i.e. becoming more magnetically dominated) up to 2~au \citep[][]{Bavassano2000}. At the orbit of the Earth, residual energy is on average negative \citep[e.g.,][]{Perri2010,Chen2013,Soljento2023}.

\citeauthor{Borovsky2020}'s (\citeyear{Borovsky2020}) statistical study of inertial range fluctuations associated with 109 shocks with density compression ratio larger than $\sim 2$ included examination of Alfv\'enicity as defined in terms of the degree and sign of correlation between the magnetic field and velocity fluctuations and  Alfv\'en ratio.  The authors found that Alfv\'enicity on average decreased at the shock transition from upstream to downstream (region durations 60--120 min) regardless of the driver of the shock. \cite{Soljento2023} analysed differences in the distributions of normalised cross-helicity and residual energy upstream and downstream of 74 ICME-driven shock waves, with their study considering the whole sheath region (up to $\sim$1 day in duration) downstream of the shock. They found that turbulence became more balanced (i.e. more $\sigma_c \sim 0$ values) downstream and that there was slightly more energy in magnetic field than in velocity fluctuations (i.e. more negative $\sigma_r$ values) in the sheath. The latter finding is consistent with the \cite{Borovsky2020} work. These findings are also consistent with the study by \cite{Good2022}, who performed a superposed epoch analysis of cross-helicity and residual energy in 176 ICME-driven sheath regions, of which 97 were bounded by shocks. Previous theoretical studies~\citep[e.g.,][]{vainio1998,vainio1999} and simulations~\citep[e.g.,][]{sishtla2023} may explain the transition from imbalanced to balanced turbulence via the presence of both transmitted (anti-sunward propagating) and reflected (sunward propagating) fluctuations as imbalanced upstream waves interact with the shock.

Most studies on the effect of shocks on normalised cross-helicity, residual energy and magnetic helicity have been case studies. We summarise in the following some of their key findings. \cite{Zhao2021} analysed an ICME-driven shock wave that was observed on 19 April 2020 by \textit{Solar Orbiter} at 0.8~au and the next day by \textit{Wind} at $\sim 1$~au. They found at \textit{Wind} that the cross-helicity at injection scales (time scales from 1~h upward) was almost zero downstream of the shock, suggesting balanced turbulence with waves propagating both parallel and anti-parallel to the magnetic field. The residual energy in turn changed at the shock transition at \textit{Wind} to a more negative value, indicating more power being in the magnetic field fluctuations. These results are consistent with the statistical studies by \cite{Borovsky2020} and \cite{Soljento2023} covering inertial range fluctuations. \cite{Zhao2021} also noted that inertial range magnetic helicity values, analysed both at \textit{Solar Orbiter} and \textit{Wind}, were enhanced both upstream and downstream of the shock and coincided with increased wave activity. Their observations suggested the presence of kinetic Alfv\'en waves at the proton cyclotron frequency downstream and lower-frequency and non-compressive ULF range waves excited by streaming particles upstream. \cite{Trotta2024a} studied a strong shock wave that was detected by \textit{Parker Solar Probe} at 0.07~au on 5 September 2022 and by \textit{Solar Orbiter} at 0.7~au the following day. Similar to \cite{Zhao2021}, this event showed enhancement of magnetic helicity close to the proton cyclotron frequency in the downstream, while the signature disappeared downstream. In a coupled turbulence transport model, \cite{Adhikari2016} identified increases in $\sigma_r$ and $|\sigma_c|$ across the shock that were in some agreement with observations from the \textit{Wind} spacecraft at 1~au, although three of the four observed shocks that they analysed in detail saw a decrease in $\sigma_r$ in the downstream and more variable $|\sigma_c|$ behaviour. 

In this work, we perform to our knowledge the first comprehensive statistical analysis of how the normalised cross-helicity, residual energy and magnetic helicity vary at interplanetary shock waves that takes into account the effect of the shock properties. Our analysis uses 371 shocks observed by the \textit{Wind} spacecraft in the near-Earth solar wind and seven shocks observed by \textit{Solar Orbiter} between 0.3\,--\,0.5~au. 

\section{Data and approaches}

\subsection{Data and event selection}
\label{sec:data_events}

We have used data from the \textit{Wind} \citep{ogilvie1997} and \textit{Solar Orbiter}  \citep{Muller2020} spacecraft. The \textit{Wind} data extends from 1995 -- 2023, a time range spanning 2.5 solar cycles, while \textit{Solar Orbiter} shocks are analysed for the period 2020 -- 2023, covering the rising phase of Solar Cycle 25. The magnetic field observations from \textit{Wind} are provided by the Magnetic Fields Investigation \citep[MFI;][]{lepping1995} and the plasma data by the Three-Dimensional Plasma and Energetic Particle Investigation \citep[3DP;][]{lin1995}. The nominal cadence of the field and plasma data used was 3~s. As the cadence varied slightly, the data were averaged to  10~s cadence. From \textit{Solar Orbiter}, we have used magnetic field measurements from the magnetometer \citep[MAG;][]{Horbury2020} and plasma data from the Solar Wind Analyser \citep[SWA;][]{Owen2020} suite. Both \textit{Wind} and \textit{Solar Orbiter} data were obtained from the NASA Goddard Space Flight Center Coordinated Data Analysis Web\footnote{\url{http://cdaweb.gsfc.nasa.gov/}} (CDAWeb). \textit{Solar Orbiter} was launched on 10 February 2020. The spacecraft has a heliocentric orbit with a perihelion distance of 0.28 au and aphelion distance of $\sim$1 au. For \textit{Solar Orbiter} the nominal cadence of the  plasma data was 4~s and for the magnetic field 0.125~s, which were also interpolated to 10~s cadence. 

The shocks were gathered from the Heliospheric Shock database\footnote{\url{http://ipshocks.helsinki.fi/}} developed and maintained at the University of Helsinki \citep[][]{Kilpua2015} as well as from the Harvard-Smithsonian Center for Astrophysics interplanetary shock database for the \textit{Wind} spacecraft\footnote{\url{https://cfa.harvard.edu/shocks/wi_data/}}. Only fast forward shocks have been analysed in this study. We excluded shocks that had data gaps in the upstream or downstream. The \textit{Solar Orbiter} shocks were obtained from the SERPENTINE project shock catalogue\footnote{\url{https://data.serpentine-h2020.eu/catalogs/shock-sc25}} \citep{Trotta2024b}. We selected only those shocks from \textit{Solar Orbiter} that were observed below 0.5 au and for which there were both plasma and magnetic field data available without data gaps in their upstream and downstream. The final set of analysed events includes 371 \textit{Wind} shocks and seven \textit{Solar Orbiter} shocks.

\subsection{Calculation of shock parameters}
\label{sec:calc_shock}

We divide the analysed shocks into different categories according to their properties that include: the gas compression ratio ($r_\mathrm{g}$), i.e. the downstream to upstream density ratio; the upstream plasma beta ($\upbeta_{\mathrm{u}}$), i.e., the ratio of plasma to magnetic pressure; the velocity jump across the shock ($\Delta V$); and the shock obliquity or shock angle ($\theta_{\mathrm{Bn}}$), i.e., the angle between the upstream magnetic field direction and shock normal. The shock angle $\theta_{\mathrm{Bn}}$ plays a fundamental role in shaping the shock structure and in  accelerating charged particles as it controls the dynamics of charged particles close to the shock. Quasi-parallel shocks \citep{Burgess2005}, which have $\theta_{\mathrm{Bn}} < 45^{\circ}$, have extended foreshock regions where particles can escape upstream from the shock and, therefore, are related to smaller and more gradual increases in the field magnitude and plasma parameters than quasi-perpendicular shocks \citep{bale2005}; quasi-perpendicular shocks, which have $\theta_{\mathrm{Bn}} > 45^{\circ}$), have particles that stay close to the shock and a sharper shock transition. The upstream plasma beta can affect the shock vicinity by controlling the generation and growth of plasma waves; for example, high plasma beta gives rise to whistler waves and enhances the amplitude of turbulent fluctuations. 

The shock parameters for all shocks have been determined using a publicly available software package, SerPyShocks \citep{Trotta2022}, which allows calculation of basic shock properties as a function of mean values calculated over varying upstream and downstream windows. We have used maximum 20-min and minimum 2-min averaging windows for both the upstream and downstream, excluding 1 min before and after the shock. To calculate the speed jumps, only the values obtained using the the maximum averaging window (20 min) both for the upstream and downstream have been used.

Shock angles have been estimated using the mixed-mode method, which incorporates both the magnetic field and velocity data in the calculation of the shock normal \citep{Abraham1976,Trotta2022}: 
$$
\hat{\mathbf{n}}_{\mathrm{sh}} = \pm \frac{(\Delta \mathbf{B} \times \Delta \mathbf{V}) \times  \Delta \mathbf{V}}{|(\Delta \mathbf{B}  \times \Delta \mathbf{V}) \times  \Delta \mathbf{V}|}
$$
where $\Delta \mathbf{B}$ and $\Delta \mathbf{V}$ are changes in the magnetic field and velocity vectors from downstream to upstream, respectively. 

\subsection{Calculation of turbulence parameters}
\label{sec:calc_turb}

The cross-helicity and residual energy can be calculated from the power spectral densities (PSDs) of the velocity and magnetic field. Here the Elsässer variables help to simplify the analysis. These variables were first defined and used by \cite{Elsasser1950} to transform the incompressible magnetohydrodynamic (MHD) equations into a symmetric form. The Elsässer variables are given by $\textbf{z}^\pm = \textbf{v} \pm \textbf{b}$, where $\textbf{v}$ represents the velocity and $\textbf{b} = \textbf{B}/\sqrt{\mu_0 \rho}$ is the magnetic field in velocity units, where $\rho$ is the ion density. Fluctuations in $\textbf{z}^-$ represent Alfv\'enic wave packets propagating parallel to the magnetic field, and fluctuations in $\textbf{z}^+$ are wave packets propagating anti-parallel to the magnetic field.  

As stated in Sect.~\ref{sec:introduction}, the residual energy is defined as the partition of energy between kinetic and magnetic fluctuations, i.e., the difference in the trace PSDs of $\textbf{v}$ and $\textbf{b}$, denoted as $E_v$ and $E_b$, respectively. The quantity is typically normalised to take values between $[-1,1]$, and is thus calculated as 
\begin{equation}\label{eq:resen}
    \sigma_r = \frac{E_v - E_b}{E_v + E_b},
\end{equation}
with negative (positive) values indicating an excess of magnetic (kinetic) fluctuation power and $\sigma_r \sim 0$ an equipartition of energy, which is a property of ideal Alfv\'en waves.  

The cross-helicity may be defined as the balance between the fluctuations propagating parallel and anti-parallel to the magnetic field, and thus is given by the difference in the trace PSD of $\textbf{z}^+$ and $\textbf{z}^-$. Similar to residual energy, the cross-helicity is also normalised to take values between $[-1,1]$:
\begin{equation}\label{eq:crossh}
    \sigma_c = \frac{E_+ - E_-}{E_+ + E_-}.
\end{equation}
Here negative (positive) values indicate greater power in fluctuations propagating parallel (anti-parallel) to the background field, and $\sigma_c \sim 0$ gives the balanced case. 

The normalised magnetic helicity \citep{Matthaeus1982,Zhao2021} is calculated as 
\begin{equation}
    \sigma_m = \frac{2 \operatorname{Im} \left[ W^{\ast}_j(\nu, t) \cdot W_k(\nu, t) \right] }{|W_i(\nu, t)|^2 + |W_j(\nu, t)|^2 + |W_k(\nu, t)|^2},
\end{equation}
where  $W_i(\nu, t)$, $W_j(\nu, t)$, and $W_k(\nu, t)$ are the wavelet transforms of the Cartesian magnetic field components $i$, $j$ and $k$ (e.g., the GSE or RTN coordinate systems) and $\nu$ is the frequency of the wavelet function. The negative values of magnetic helicity indicated left-handed polarised waves while positive values indicate right-handed polarised waves. In the solar wind at $\sim$\,1 au, normalised magnetic helicity in the injection range is on average strongly negative, on average zero in the inertial range and positive in the dissipation range \citep[e.g.,][]{Smith2003}.

We have calculated $\sigma_c$, $\sigma_r$ and $\sigma_m$ in 1-h regions both upstream and downstream of the shocks, excluding 1-min intervals immediately before and after the shock time. The wavelet spectrograms were calculated separately for these intervals, and values outside the cone of influence have been removed. Probability distribution functions (PDFs) have been built for $\sigma_c$, $\sigma_r$ and $\sigma_m$ for all shocks, as well as for shock subsets separated by the shock parameters. 

The direction of the interplanetary magnetic field (IMF), which defines the IMF sector, can be either towards or away from the Sun. Thus whether fluctuations in $\textbf{z}^\pm$ correspond to sunward or anti-sunward fluctuations is dependent on the IMF sector. To explore in more detail how the shock transition affects the relative power in sunward and anti-sunward propagating waves, we also find the rectified cross-helicity, $\sigma_c^*$, where $\textbf{z}^+$ ($\textbf{z}^-$) fluctuations are fixed to always corresponds to anti-sunward (sunward) fluctuations, and positive (negative) $\sigma_c^*$ indicates an anti-sunward (sunward) imbalance. The winding angle of the Parker spiral in the ecliptic plane, $\phi_{\textrm{Parker}}$, is determined by the radial solar wind speed, $v_{r,SW}$, radial distance from the Sun, $r$, and the angular velocity of the Sun, $\Omega$, such that $\phi_{\textrm{Parker}} = \tan^{-1}((r \Omega) / v_{r,SW})$. For data in GSE coordinates at 1 au, the \textit{towards} sector has on average an IMF clock angle $\phi$ below $45^{\circ}$ or over $225^{\circ}$, while in the \textit{away} sector $\phi$ is in the interval $[45^{\circ},225^{\circ}]$; the Parker spiral winding angles at 1~au are thus on average $315^{\circ}$ and $135^{\circ}$ in the towards and away sectors, respectively.

To perform the rectification, the magnetic field sign is flipped when the IMF is in the away sector before calculating the cross-helicity. For this analysis, we only use cases where the IMF stays consistently in one sector throughout the 2-h interval investigated to avoid inconsistencies from IMF sector reversals. This part of the analysis was only performed for the \textit{Wind} shocks due to the relatively low number of \textit{Solar Orbiter} shocks available. In total, 233 \textit{Wind} shocks were investigated for their effects on $\sigma_c^*$. 

\section{Results}
\label{sec:results}

\subsection{Example events}
\label{sec:examples}

\begin{figure}[ht]
\centering
\includegraphics[width=\linewidth]{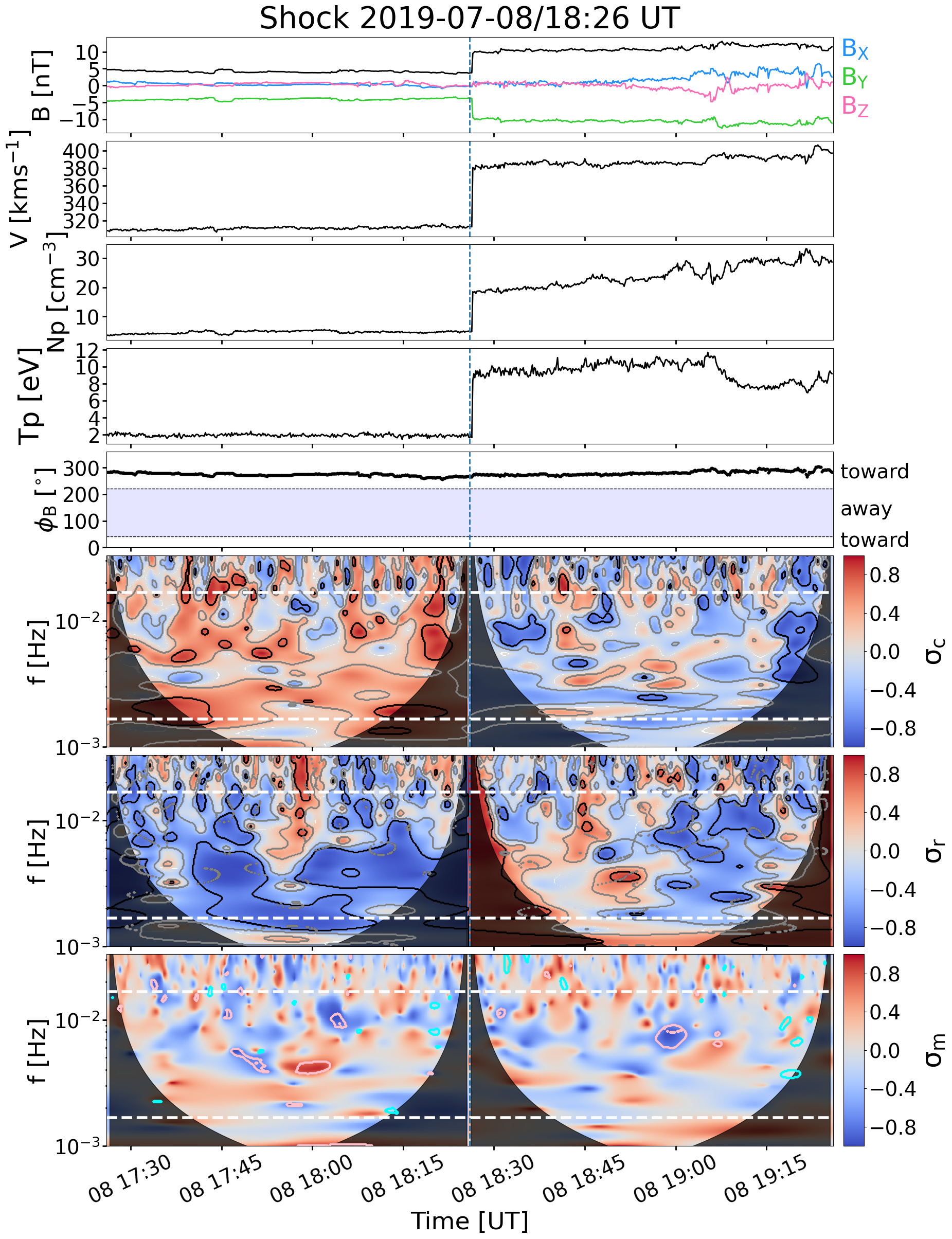}
\caption{A shock observed by \textit{Wind} on 8 July 2019. From top to bottom, the first five panels show: the magnetic field magnitude and the GSE components; solar wind speed; density; temperature; and IMF clock angle. The bottom three panels show wavelet spectrograms of normalised cross-helicity, residual energy and magnetic helicity. Dashed white lines show the frequencies limiting the high frequencies / small scales (16.7 mHz / 1 min) and low frequencies / large scales (1.67 mHz /10 min) used in the analysis. Black contours in the spectrograms delineate where absolute values of parameters exceed 0.7, and grey contours outline $|\sigma_r| > 0.3$ and $|\sigma_c| > 0.3$ regions. The pink and cyan contours in the bottom panel outline the intervals that fulfil the criteria for Alfv\'enic fluctuations and flux ropes imposed on $\sigma_c$, $\sigma_r$, and $\sigma_m$ (see Sect.~\ref{sec:change_awfr} for details).}
\label{fig:example1}
\end{figure}

Here we describe a few examples from the full set of analysed events. Figure~\ref{fig:example1} shows a shock detected at \textit{Wind} on 8 July 2019 at 18:26 UT. The shock was nearly perpendicular, with $\theta_{Bn} = 83.9^{\circ}$, moderate values of $\Delta V=45.5\,\textrm{km\,s}^{-1}$ and $r_\mathrm{g}=2.0$, and $\upbeta_{\mathrm{u}}=2.1$. The IMF cone angle was consistently in the towards sector during the 2-hr period shown and thus positive $\sigma_c$ values (red) indicate anti-sunward propagation and negative $\sigma_c$ values (blue) sunward propagation. The upstream region was dominated by anti-sunward fluctuations at all frequencies, but there were some local patches of predominantly sunward fluctuations and fluctuations with $|\sigma_c| \sim 0$. The turbulence in the upstream is however globally balanced \citep[][]{Chen2013}, indicated by the average value of $|\sigma_c|$ being 0.20. The fluctuations in the downstream region are also relatively balanced with the average $|\sigma_c| = 0.28$ but now, in contrast to the upstream, sunward fluctuations dominate. The residual energy was mostly negative in the upstream with the average value of $\langle \sigma_r \rangle = -0.48$, indicating a clear excess of magnetic fluctuation energy, while in the downstream the average is $\langle \sigma_r \rangle = -0.19$ showing global  equipartitioning. The downstream region closest to the shock exhibits patches with more power in kinetic fluctuations whereas the region deeper in the downstream is dominated by patches with more power in magnetic fluctuations. The magnetic helicity in turn did not show clear changes from the upstream to downstream, having values mostly quite close to zero as is usually the case in the inertial range for solar wind at 1~au (see Sect.~\ref{sec:calc_turb}). The average magnetic helicities in the upstream and downstream are $-0.024$ and $-0.073$, respectively. 

\begin{figure}[ht]
\centering
\includegraphics[width=\linewidth]{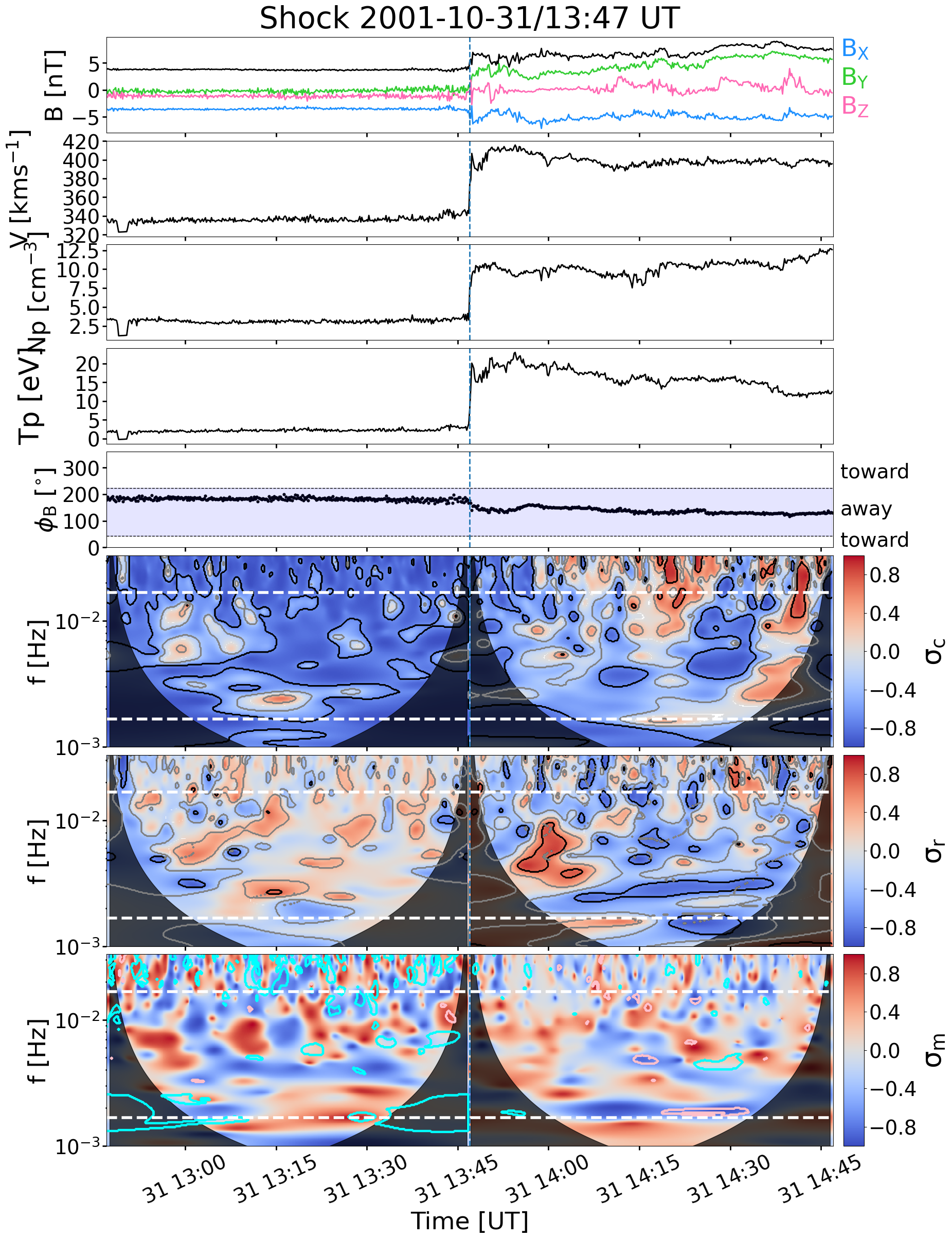}
\caption{A shock observed by \textit{Wind} on 31 October 2001. The panel layout is the same as in Fig.~\ref{fig:example1}.}
\label{fig:example2}
\end{figure}

Another example of a shock detected by \textit{Wind} is shown in Fig.~\ref{fig:example2}, which occurred on 31 October 2001 at 13:47 UT. It was nearly parallel, with $\theta_{Bn} = 9.69^{\circ}$, $\Delta V=68.7 \,\textrm{km\,s}^{-1}$, $r_\mathrm{g}=3.3$, and $\upbeta_{\mathrm{u}}=1.4$. The IMF was in the away sector and so anti-sunward fluctuations had negative (blue) and sunward fluctuations positive (red) cross-helicity values. The majority of fluctuation power in both the upstream and downstream was anti-sunward. The upstream is clearly unbalanced with average $|\sigma_c| = 0.64$ while in the downstream cross-helicity is again relatively balanced with average $|\sigma_c| = 0.30$. The residual energy showed significant equipartition in the upstream region (average $\sigma_r = -0.027$) with patches of negative and positive values. The downstream region was instead characterised by predominantly negative residual energy ($\sigma_r = -0.27$), except for a localised region just after the shock. Both the upstream and downstream regions had significant instances of strong magnetic helicity at all frequencies, particularly in the upstream.  The average $\sigma_m$ values were however again close to zero with values $-0.059$ and $-0.036$ in the upstream and downstream, respectively. 

The cyan and pink contours in the bottom panels of Figs.~\ref{fig:example1} and \ref{fig:example2} delineate regions fulfilling the criteria for Alfv\'{e}n waves and small-scale flux ropes; this aspect of the analysis is discussed in Sect.~\ref{sec:shock_turbulence}.

\subsection{Shock parameters}
\label{sec:shocks}

\begin{figure*}
\captionsetup[subfigure]{labelformat=empty}
\centering
   \begin{subfigure}[b]{0.99\linewidth}
   \includegraphics[width=0.99\linewidth]{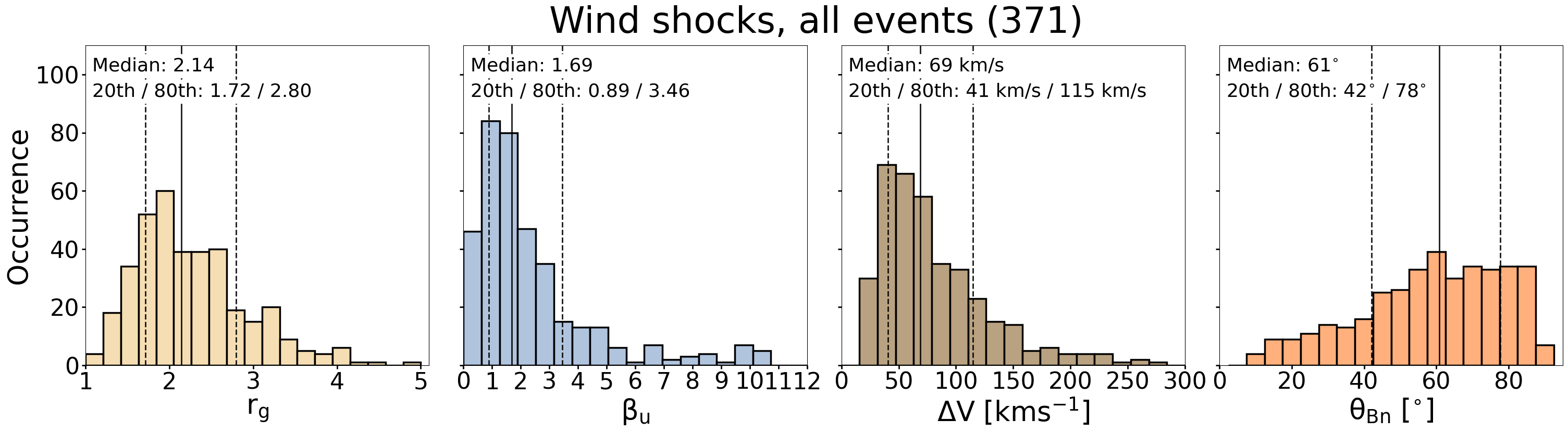}
   \caption{}
\end{subfigure}

\begin{subfigure}[b]{0.99\linewidth}
   \includegraphics[width=0.99\linewidth]{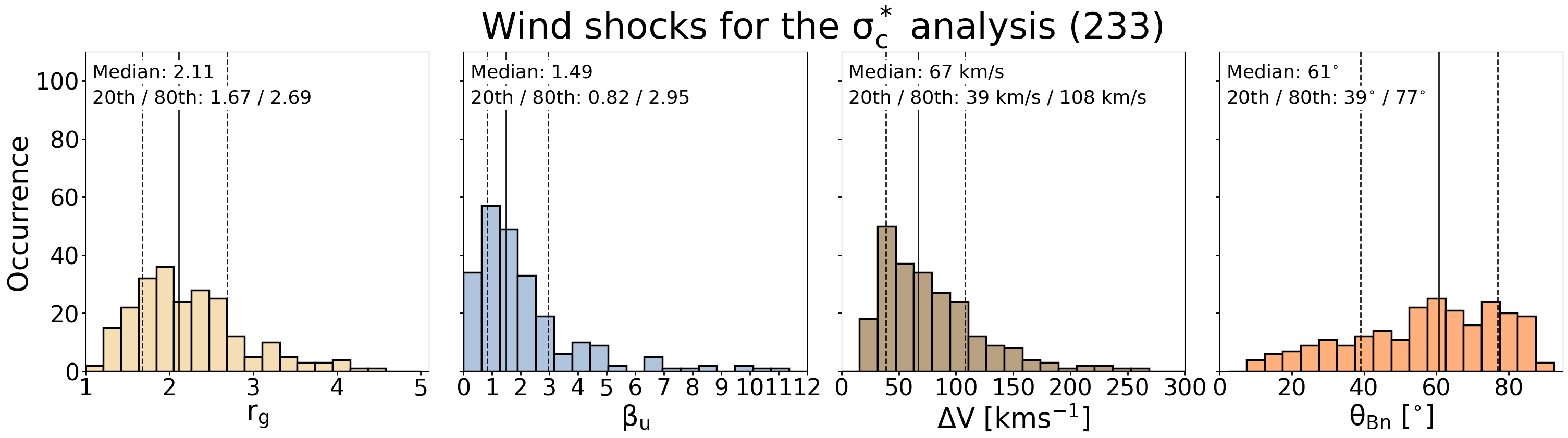}
   \caption{}
\end{subfigure}
\caption{Histograms of parameters for interplanetary shocks detected by the \textit{Wind} spacecraft during 1995--2023. The dashed grey lines show the 20th and 80th percentiles. The top row shows the distribution for all \textit{Wind} shocks and the bottom row shows the distribution for those used in the $\sigma_c^*$ analysis. From left to right, the panels show the distributions for the shock gas compression ratio, upstream plasma beta, velocity jump across the shock and shock angle.}
   \label{fig:shock_distribution}
\end{figure*}

Figure~\ref{fig:shock_distribution} shows histograms of the selected shock parameters, namely the gas compression ratio (i.e. downstream-to-upstream density ratio), plasma beta, velocity jump and shock angle across all 371 shocks from the \textit{Wind} spacecraft, and also for the 233 shocks for which the rectified cross-helicity ($\sigma_c^*$; Sect.~\ref{sec:calc_turb}) could be determined. The solid vertical lines indicate the medians and the dashed lines give the 20th and 80th percentiles that we later use to divide the shocks into two subsets. These percentiles were chosen to give sufficiently distinct populations but with a sufficiently large number of events (74 in each of the subsets) for the robust statistical analysis. 

The median values of $r_\mathrm{g}$ are slightly over 2.0 for all events and for those used in the $\sigma_c^*$ analysis. The histograms are biased towards small $r_\mathrm{g}$ values but have tails that extend to $r_\mathrm{g}>4$, i.e., beyond the theoretical limit for perpendicular MHD shocks. For the majority of shocks, $\upbeta_{\mathrm{u}} > 1$, with the distributions having long $\upbeta_{\mathrm{u}}$ tails. The 20th percentile of $\upbeta_{\mathrm{u}}$, however, is at 0.89; the 20th percentile subset thus represents cases where the magnetic pressure dominates in the upstream. The upper quartile population with  $\upbeta_{\mathrm{u}}  \gtrsim 3$ represent in turn cases that have their upstream beta clearly above the typical solar wind values, 
\citep[$\upbeta_{\mathrm{u}}  \sim 1-2$; e.g.,][]{Keyser2001,Mullan2006}.

The $\Delta V$ distribution peaks at $\sim 70 \,\textrm{km\,s}^{-1}$ with a tail extending to $\Delta V \sim  250\,\textrm{km\,s}^{-1}$. Finally, the $\theta_{Bn}$ angles indicate a clear preponderance of quasi-perpendicular shocks ($\thetaup_\mathrm{Bn} > 45^{\circ}$) with the median being $61^{\circ}$. As discussed in \cite{Kilpua2015}, this could be partly a selection bias as quasi-parallel shocks are more difficult to identify from the ambient wind due to their tendency for more gradual transitions from upstream to downstream, complex structure of the shock surroundings, and more modest field jumps. However, the shocks in the 20th percentile subset are all in the quasi-parallel regime ($\thetaup_\mathrm{Bn} < 45^{\circ}$).

\subsection{Statistical results at 1 au}
\label{sec:statistical}

\subsubsection{Shock influence on turbulence parameters}
\label{sec:shock_turbulence}

\begin{figure*}[t]
\includegraphics[width=0.99\linewidth]{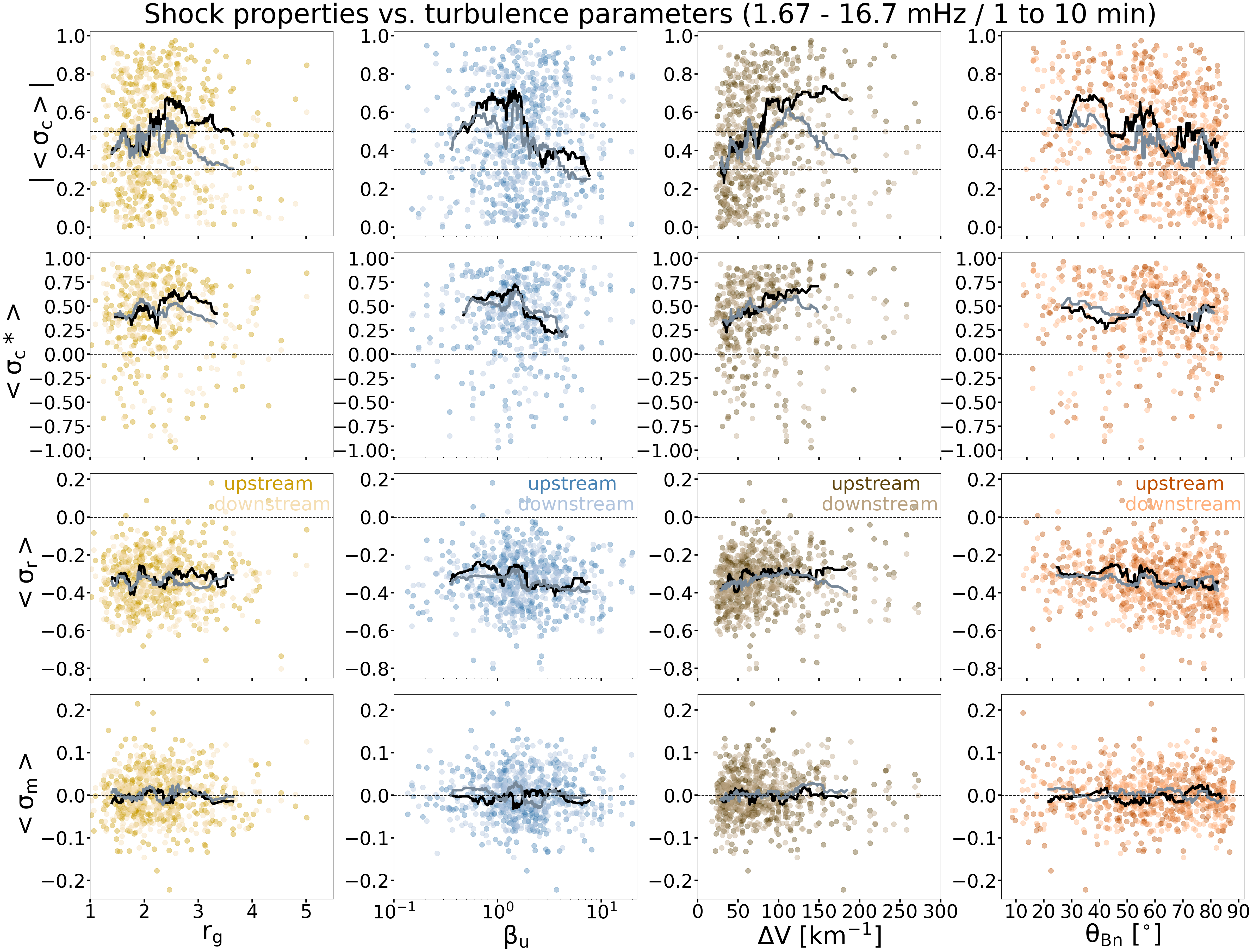}
\caption{Turbulence parameters as a function of shock properties. From top to bottom, the rows show the absolute value of the 1-h average of cross-helicity, 1-h average of the rectified cross-helicity, 1-h average of residual energy and 1-h average of magnetic helicity. The values are shown both for the upstream (darker shades) and downstream (paler shades) as a function (from left to right) of shock gas compression ratio, upstream plasma beta, velocity jump at the shock and shock angle. Curves give the 40-event running median for the upstream (black) and downstream (grey). The x-axis ranges have been limited to exclude a few outlier points to enhance visibility of the median curves; the full ranges are shown in Supplementary Figure A1.}
   \label{fig:shock_turbulence}
\end{figure*}

We first investigate how the turbulence parameters in the upstream and downstream depend on the selected shock parameters. The results are shown in Fig.~\ref{fig:shock_turbulence} for the individual events in the upstream (darker colours) and downstream (lighter colours). The values are calculated from the wavelet spectrograms over frequencies 1.67-16.7 mHz (1-10 min timescales), which fall within the inertial range of MHD turbulence \citep[e.g.,][]{Bruno2013,Verscharen2019} at the orbit of the Earth. The curves in Fig.~\ref{fig:shock_turbulence} give 40-event running medians.

The points in the two top panels show the absolute values of the 1-h averages of normalised cross-helicity ($|\langle \sigma_c \rangle|$) and 1-h averages of the rectified cross-helicity ($\langle \sigma_c^* \rangle$). We have chosen to show the absolute values of the averages for the cross-helicity to give a better estimate of how balanced or imbalanced the turbulence is. This is because the balance and imbalance are global rather than local properties \citep[e.g.,][]{Chen2013}, and taking the magnitude first would lose information about the global signed average. As detailed in Sect.~\ref{sec:calc_turb}, the rectified cross-helicity fixes the propagation direction with respect to the Sun; negative rectified cross-helicity indicates sunward propagation and positive values represent anti-sunward propagation in the plasma frame. The two bottom panels show the averages of the residual energy and magnetic helicity ($\langle \sigma_r \rangle$ and $ \langle \sigma_m  \rangle$, respectively). 

Before discussing the dependence on the shock characteristics, we will summarise some overall properties of the investigated turbulence parameters. 
Firstly, the two top panels of Fig.~\ref{fig:shock_turbulence} show that $|\langle \sigma_c \rangle|$ and $\langle \sigma_c^* \rangle$ spread over all possible values. The means over the whole data set are shown in Table \ref{table:means} as well as the percentage of the events that had $|\langle \sigma_c \rangle| < 0.3$, i.e. globally balanced fluctuations, and $|\langle \sigma_c \rangle| > 0.5$, i.e. globally imbalanced fluctuations. Both in the upstream and downstream about one-third of the events have balanced fluctuations while the upstream has considerably more imbalanced events, 50\% compared to 40\%, respectively. The fluctuations propagate predominantly anti-sunward as featured by the clear majority of $\langle \sigma_c^* \rangle$ values being positive (over 80\%, Table \ref{table:means}). The residual energies are in turn nearly all negative indicating that magnetic energy in fluctuations dominates over kinetic energy. According to Table \ref{table:means} the percentage of the events where fluctuations are close to global equipartition, i.e., $|\langle \sigma_r  \rangle | < 0.3$, is considerably higher for the upstream than for the downstream (46\% and 36\%, respectively). Finally,  Fig.~\ref{fig:shock_turbulence} shows that the average magnetic helicities are clustered around zero.

\begin{table}[!h] \caption{Mean values of key parameters and percentages of time when fluctuations are balanced, imbalanced, predominantly anti-sunward, and equipartitioned, and when Alfv\'enic fluctuations (AFs) and small-scale flux ropes (SFRs) are present.}
\centering
\begin{tabular}{c c c}
\hline\hline
  & upstream &  downstream \\
  \hline
mean $|\langle \sigma_c \rangle|$   & 0.49 & 0.43 \\
mean $\langle \sigma_c^* \rangle$   & 0.38 & 0.36 \\
mean $\langle  \sigma_r  \rangle$   & -0.33 & -0.35 \\
mean $\langle  \sigma_m  \rangle$  & $\sim 0$ & $\sim 0$ \\
balanced ($|\langle \sigma_c \rangle| < 0.3$)  &   31\%    &  32\% \\
imbalanced ($|\langle \sigma_c \rangle| > 0.5$)  &   50\%   &  40\% \\
anti-sunward ($\langle \sigma_c^* \rangle > 0$)  &   82\%     &  84\% \\
equipartitioned ($|\langle \sigma_r| \rangle| < 0.3$)  &   46\%     &  36\% \\
mean AF occurrence & 14.8\% & 10.8 \%  \\
mean SFR occurrence & 1.1\% & 1.7 \% \\
\hline 
\end{tabular} \label{table:means}
\end{table}

Figure~\ref{fig:shock_turbulence} shows a large scatter as a function of the displayed shocks parameters for all cases but some trends are visible. The $|\langle \sigma_c \rangle |$ values (also the rectified ones) are the highest, i.e., feature the highest imbalance, for $\textrm{r}_\textrm{g} \sim 2.5$ after which the median curves clearly decline towards more balanced fluctuations. The second columns in the top rows show that the cross-helicities both in the upstream and downstream exhibit the highest imbalance of fluctuations for $\upbeta_{\mathrm{u}}  \sim 1-2$, which are, as previously mentioned,  common 1 au values. For 
 the lowest ($\upbeta_{\mathrm{u}}  \lesssim 1$) and higher  ($\upbeta_{\mathrm{u}}  \gtrsim 3$) upstream beta values a considerably smaller fraction of fluctuations are imbalanced (except for the very largest plasma beta, but the number of events too small to draw strong conclusions). Both the upstream and downstream also exhibit clear increase of $| \langle \sigma_c \rangle |$ and $\langle \sigma_c^* \rangle$ with the increasing shock velocity jump up to $\Delta \textrm{V} \sim 100 - 150\,\textrm{km\,s}^{-1}$ after which in the upstream the curves level off while in the downstream fluctuations become again more balanced.  Finally, the last panel in the top row shows that there is a weak tendency for fluctuations to be more balanced for quasi-perpendicular shocks than for quasi-parallel shocks, but the median curves show large fluctuations and there are relatively few parallel shocks in the distribution. 

The $\langle \sigma_r \rangle$ and $\langle \sigma_m \rangle$ values do not show such obvious trends with the shock parameters as was previously found for the cross-helicities. The residual energy values exhibit a weak trend towards zero (fluctuations becoming more equipartitioned) with increasing $\Delta \textrm{V}$ and towards increasingly negative values with the increasing shock angle. In addition, similar to the most imbalanced fluctuations, the highest equipartitioning occurs for the upstream plasma beta in the range $\sim 1-2$. 
  
\subsubsection{Change of turbulence parameters at the shock}
\label{sec:change_shock_turbulence}

\begin{figure*}[t]
\includegraphics[width=0.99\linewidth]{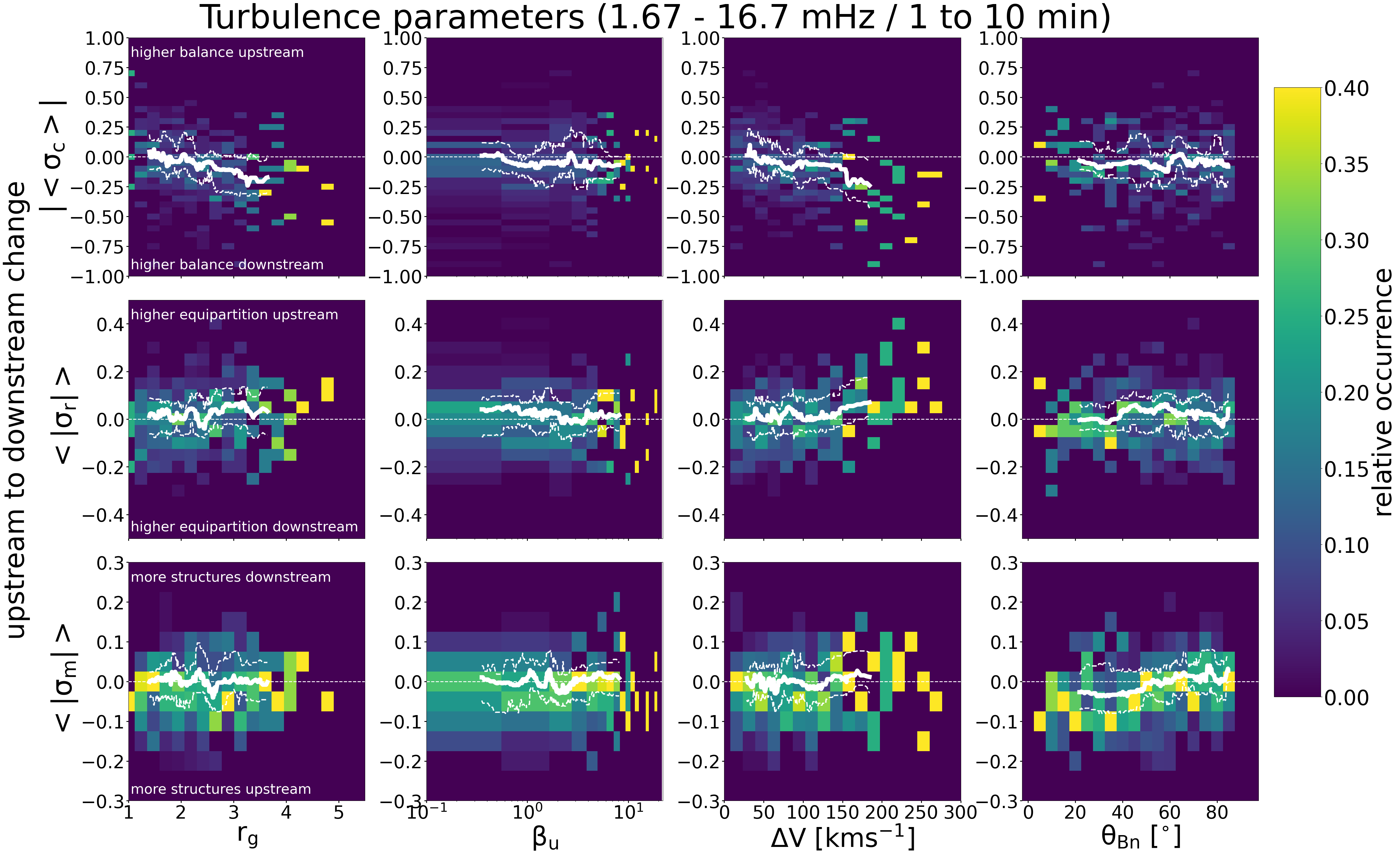}
\caption{Heatmaps of relative occurrence organised by the shock parameters, and change from upstream to downstream in (from top to bottom) the absolute value of 1-h averaged cross-helicity, 1-h averaged absolute values of residual energy and 1-h averaged absolute values of magnetic helicity. White curves give the 40-event running medians of the change. The dashed white curves give the corresponding upper and lower quartiles. The x-axis ranges have been limited to exclude a few outlier points to enhance visibility of the median curves; the full ranges are shown in Supplementary Figure A2.}
   \label{fig:change}
\end{figure*}

\begin{figure*}[t]
\includegraphics[width=0.75\linewidth]{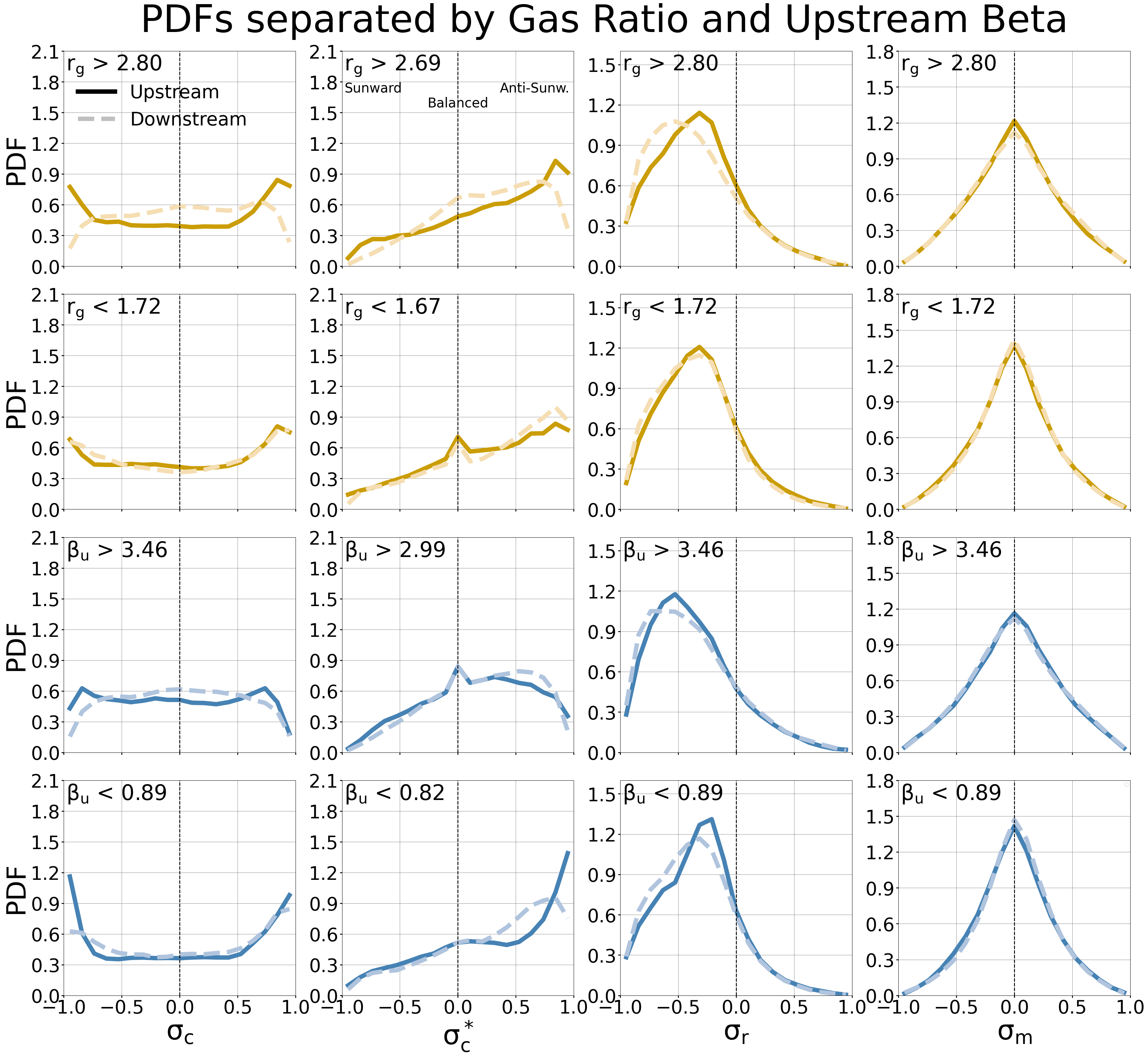}
\caption{PDFs for shocks separated into two subsets using the 80th and 20th  quartiles of the shock gas compression ratio and upstream plasma beta for cross-helicity, rectified cross-helicity, residual energy and magnetic helicity.}
   \label{fig:change_RB}
\end{figure*}

\begin{figure*}[t]
\includegraphics[width=0.75\linewidth]{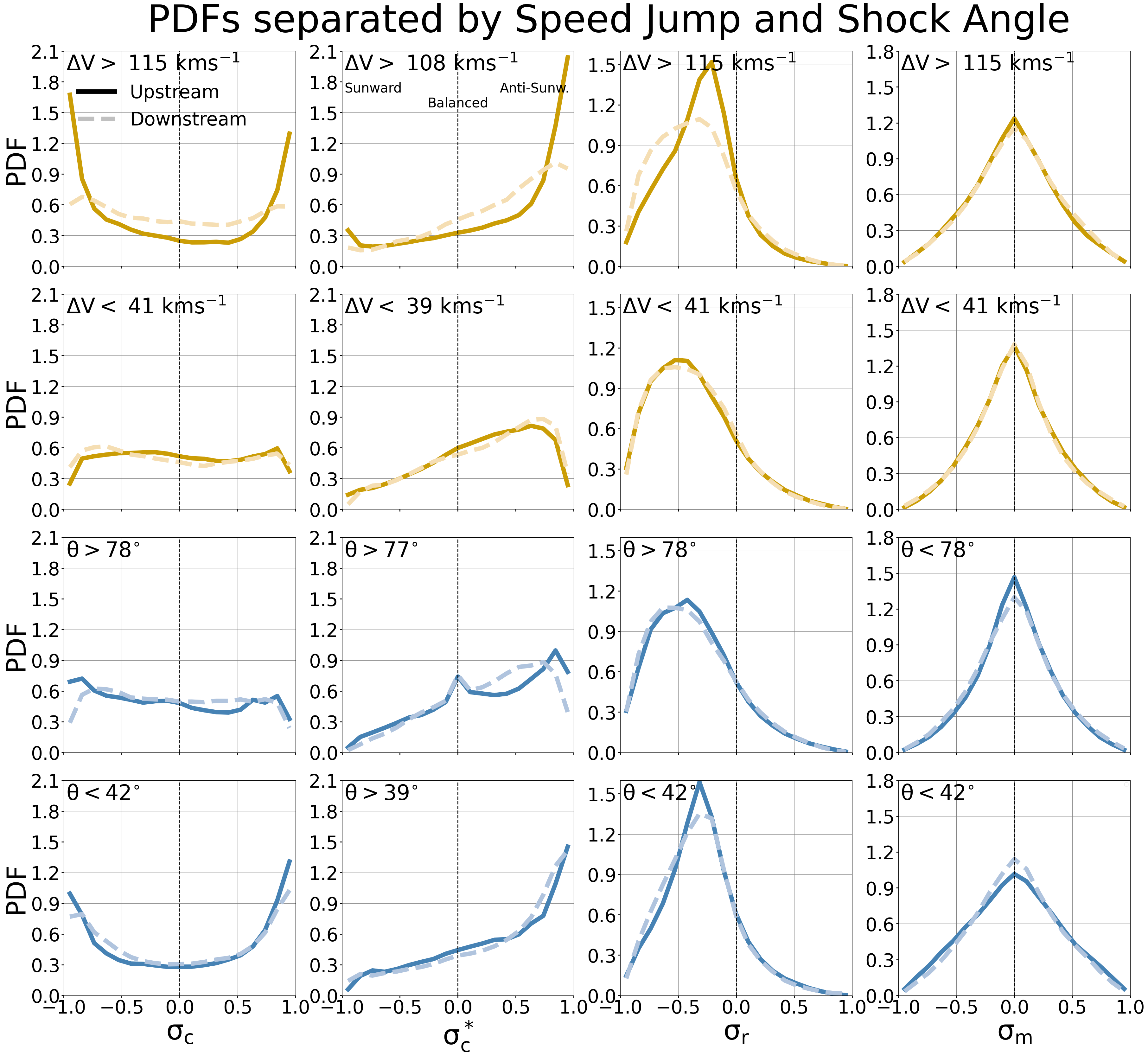}
\caption{PDFs for shocks separated into two subsets using the 80th and 20th quartiles of the shock speed jump and shock angle for cross-helicity, rectified cross-helicity, residual energy and magnetic helicity.}
   \label{fig:change_VT}
\end{figure*}

\begin{figure*}[t]
\includegraphics[width=0.99\linewidth]{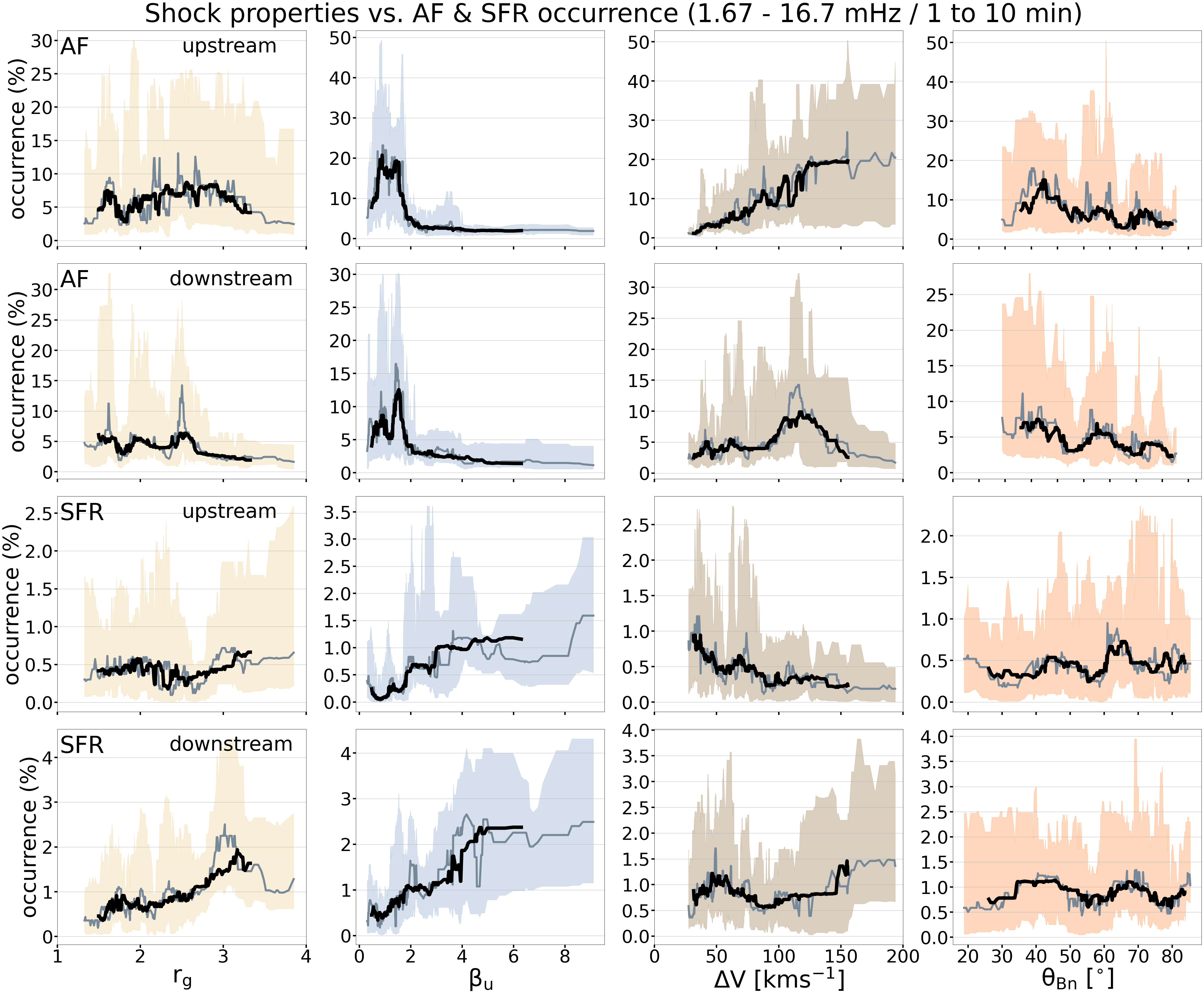}
\caption{The 30-event (grey) and 60-event (black) running medians of the occurrence percentage of periods that fulfil the AF and FR criteria in the shock upstream and downstream as functions of the selected shock parameters. Shading shows the interquartile range for the 30-event running medians.}
   \label{fig:shock_FRAC}
\end{figure*}

\begin{figure*}[t]
\includegraphics[width=0.99\linewidth]{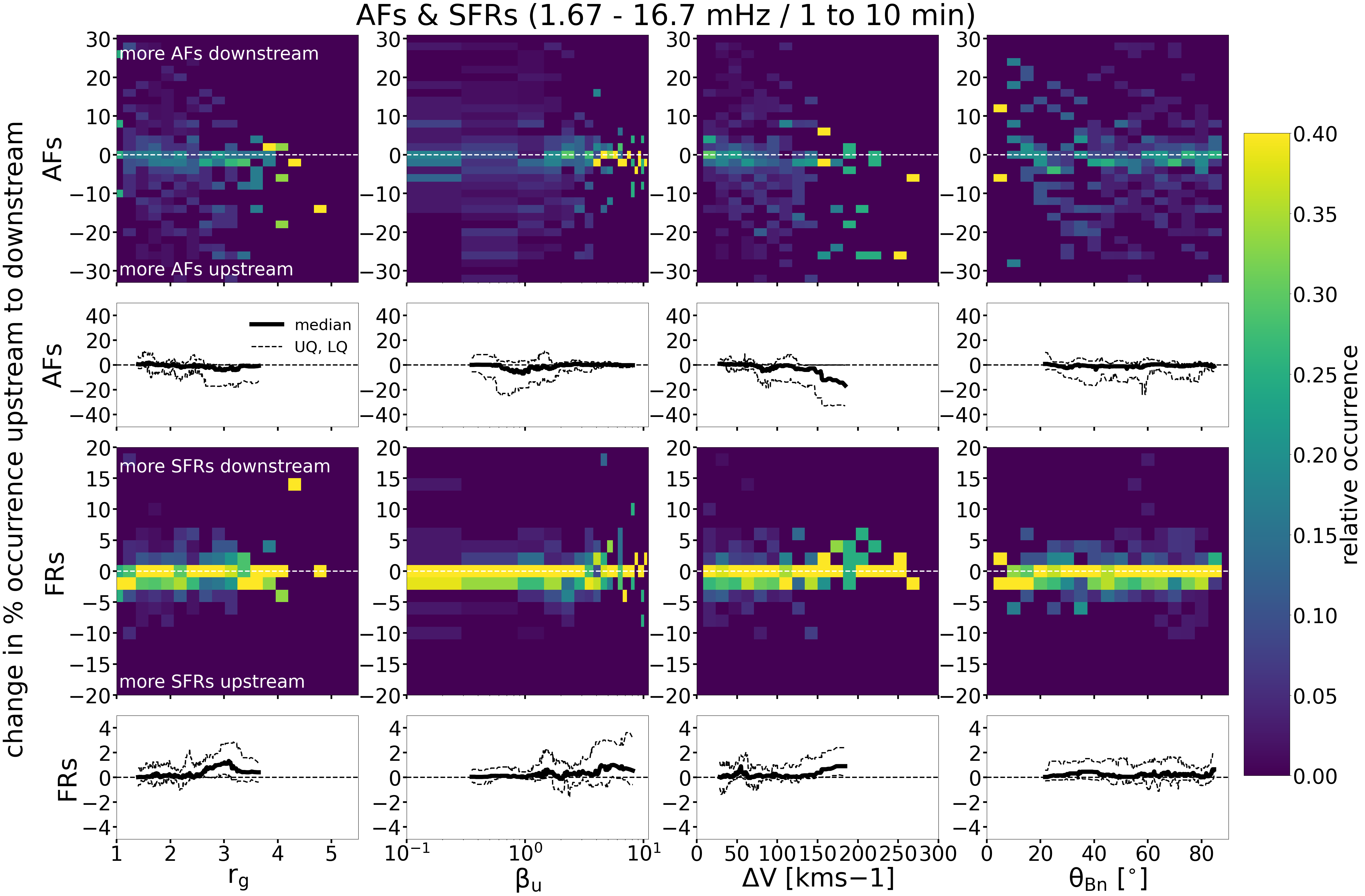}
\caption{Heatmaps of the relative occurrence rates as a function of the shock parameters and change across the shock in the percentage of AF and FR periods. The white curves give the 40-event running medians and upper and lower quartiles of the change.}
    \label{fig:change_FRAC}
\end{figure*}

We will investigate next how the average values of normalised residual energy, cross-helicity and magnetic helicity change across the shock as a function of the shock properties. The heatmaps in Fig.~\ref{fig:change} show relative occurrences in bins defined by the values of the shock parameters and change from the upstream to downstream (i.e., the upstream value subtracted from the downstream value) of $|\langle \sigma_c  \rangle|$, $\langle |\sigma_r|  \rangle$, and $\langle |\sigma_m|  \rangle$. Relative occurrences have been calculated by dividing the number of events in a given bin with the total number of events in the corresponding shock parameter range. This approach was chosen as the number of events varies considerably with the shock parameters, as shown in Figs.~\ref{fig:shock_distribution} and \ref{fig:shock_turbulence}. The white solid curves give the 40-event running medians of the percentage differences and the dashed curves the upper and lower quartiles. 

In the top row, positive (negative) percentages mean that the upstream (downstream) had more balanced $\sigma_c$ and in the middle row that the upstream (downstream) kinetic and magnetic energies were closer to equipartition. In the bottom panels, large positive (negative) values indicate that the downstream (upstream) has potentially more coherent structures, either Alfv\'{e}n waves or small-scale flux ropes. Since the $\upbeta_{\mathrm{u}}$ had a long tail to high values with only a few events we have limited the range in the plot for clarity.  

Firstly, Fig.~\ref{fig:change} shows that the upstream to downstream change in $\langle |\sigma_r|  \rangle$ and $ \langle |\sigma_m|  \rangle$ values between upstream and downstream are generally much smaller than for $| \langle \sigma_c \rangle |$ (note different y-scales in top row and other two rows). 

The top row of Fig.~\ref{fig:change} shows that for the individual events the fluctuations in the downstream are typically more balanced compared to the upstream (i.e., the change from the upstream to downstream is negative). However, there is a considerable fraction of events for which the upstream is more balanced than the downstream (39\% from total 371 events). The tendency for the downstream to be more balanced than the upstream increases with increasing shock gas compression ratio and shock velocity jump. In particular, for $\textrm{r}_\textrm{g} \gtrsim 2.5$ and $\Delta$V $\gtrsim 150$ kms$^{-1}$ the change is negative for the majority of the cases.  We also note that in our example events in Sect.~\ref{sec:examples} the first case with smaller $\textrm{r}_\textrm{g}$  had more balanced upstream while the second event with larger $\textrm{r}_\textrm{g}$ had imbalanced upstream and balanced downstream. For the shock angle and upstream plasma beta, no clear trend is visible. 

The middle panels in Fig.~\ref{fig:change} show that the upstream to downstream change in $\langle |\sigma_r| \rangle$ is slightly more frequently positive than negative signifying that kinetic and magnetic energies of fluctuations tend to be in closer equipartition in the upstream than in the downstream. This is particularly true for shocks associated with large velocity jumps ($\gtrsim 150$ kms$^{-1}$). For $\langle |\sigma_m|  \rangle$ in turn the only notable (weak) trend is that the values tend to be negative for quasi-parallel shocks which could indicate the more frequent presence of high magnetic helicity structures in the upstream compared to downstream. We examine this in more detail in Sect.~\ref{sec:change_awfr}. 

To further explore changes related to the shock properties, we present in Figs.~\ref{fig:change_RB} and \ref{fig:change_VT} how PDFs of $\sigma_c$, $\sigma_r$ and $\sigma_m$ vary from the upstream to downstream. The PDFs combine all values from the wavelet spectrograms in the 1.67mHz -- 16.7 mHz frequency range (1--10 min timescale range) in shock subgroups based on the 20th and 80th percentiles of the selected shock parameters (see Fig.~\ref{fig:shock_distribution}). 

We first note that the PDFs for magnetic helicity $\sigma_m$ are nearly symmetric around $\sigma_m =0$, consistent with previous studies finding that the magnetic helicity averages to zero in the inertial range in the solar wind.  Secondly, their PDFs show very little difference between the upstream and downstream for all investigated cases. Therefore, in the following, we discuss only variation in the PDFs for the cross-helicity and residual energy.

The top two rows in Fig.~\ref{fig:change_RB} show how the PDFs change depending on the shock gas compression ratio. The  top rows give the PDFs for the subset with gas compression ratios within the 80th percentile ($\textrm{r}_\textrm{g} > 2.8$, and $\textrm{r}_\textrm{g} > 2.69$ for $\sigma_c^*$) while the next row gives the PDFs for compression ratios within the 20th percentile ($\textrm{r}_\textrm{g} < 1.72$, and $\textrm{r}_\textrm{g} < 1.67$ for $\sigma_c^*$). The cross-helicity PDFs are almost identical upstream and downstream for the shocks with small $\textrm{r}_\textrm{g}$, while for the 80th percentile subgroup significant changes occur, as follows: in agreement with Fig.~\ref{fig:change}, the cross-helicity PDFs show that fluctuations on average are clearly more balanced downstream than in the upstream at large $\textrm{r}_\textrm{g}$, with a strong decrease in $\sigma_c \sim \pm 1$ values and enhancement in values between $\sigma_c  = -0.5$ and 0.5. The rectified cross-helicity ($\sigma_c^*$) reveals that both in the upstream and downstream the waves propagate predominantly anti-sunward. It also reveals that for the upper quartile population, both the sunward ($\sigma_c^* \sim -1$) and anti-sunward ($\sigma_c^* \sim 1$) cross-helicities decrease from upstream to downstream, with a stronger decrease observed for the anti-sunward waves.
The enhancement from upstream to downstream occurs around $\sigma_c^* \sim 0$, indicating that fluctuations become more balanced. 

The peak of the residual energy PDF for the $\textrm{r}_\textrm{g}$ upper quartile population in turn shifts to considerably more negative values and becomes slightly flatter from upstream to downstream, indicating that fluctuations in the downstream have increasingly more power in magnetic than kinetic fluctuations. For the lower quartile population in turn, the differences between the upstream and downstream PDFs are minimal. 

The two bottom rows of Fig.~\ref{fig:change_RB} show the PDFs for the 80th and 20th upstream plasma beta percentiles. The plasma beta has relatively little effect on how the combined PDFs change at the shock. There are, however, some interesting differences between the PDFs for the high and low beta cases that were not so evident from the previously shown averaged values. Firstly, the cross-helicities are considerably more balanced for high $\upbeta_u$ shocks than for low $\upbeta_u$ shocks. For high $\upbeta_u$ shocks the rectified cross-helicity PDFs peak at $\sigma_c^* \sim 0$ while for low $\upbeta_u$ the anti-sunward waves clearly dominate the distribution. In addition, the residual energy PDFs for low $\upbeta_u$ shocks peak close to zero (indicating equipartition in magnetic and kinetic power), while for high $\upbeta_u$ shocks the distribution is biased at negative values.

The top two rows in Fig.~\ref{fig:change_VT} show PDFs for the 80th and 20th percentile velocity-jump subsets. The shocks with small velocity jumps have almost identical PDFs in the upstream and downstream, but the cross-helicity and residual energy PDFs again differ considerably for shocks with large $\Delta$V: In the upstream, fluctuations are highly imbalanced with strong bias towards waves propagating predominantly anti-sunward ($\sigma_c^* \sim 1$) but there is also a population of sunward  waves ($\sigma_c^* \sim -1$) not visible for the other investigated subsets. In the shock transition, the anti-sunward population decreases strongly and turbulence becomes more balanced downstream. Similar to what was observed to $\textrm{r}_\textrm{g}$ upper quartile subset, the residual energy PDF flattens and the peak shifts towards more negative values from upstream to downstream. For shocks with small speed jumps the residual energy values are also peaked towards negative values both upstream and downstream, resembling the downstream PDF for high $\Delta$V.

For the shock angle, the differences in PDFs are relatively small between the upstream and downstream, both for the quasi-parallel and quasi-perpendicular shocks. A key difference is that the PDFs show turbulence becoming less imbalanced downstream for quasi-perpendicular shocks. Again, there are some noteworthy differences between the two subsets. The cross-helicities are more imbalanced and residual energies show higher equipartition for parallel shocks than for perpendicular shocks.

\subsubsection{Alfv\'enic fluctuations and small-scale flux ropes at the shock}
\label{sec:change_awfr}

The investigated turbulence parameters provide a means to identify small-scale flux ropes (SFRs) and Alfv\'enic fluctuations (AFs) from the solar wind plasma and magnetic field measurements. We adopt the same approach and criteria used by \cite{Zhao2021} and \cite{Ruohotie2022}, who investigated SFR and AF occurrence using \textit{Parker Solar Probe} and \textit{Wind} data.

For SFRs, we require that they exhibit a large magnetic helicity, with $|\sigma_m| > 0.7$. For AFs, we require that $|\sigma_m| > 0.7$ or $|\sigma_m| > 0.3$ for circular or linearly polarized waves, respectively. By definition, AFs have significant v-B correlations or anti-correlations, and so the criterion $|\sigma_c| > 0.9$ is imposed. Flux ropes in turn are known to have low cross-helicity  (i.e. the absence of the v-B correlations or anti-correlations of AFs) and so $|\sigma_c| < 0.4$ is required. Idealised Alfv\'enic fluctuations have equipartition of energy between magnetic and kinetic fluctuations, and $|\sigma_r| < 0.3$ is thus required for AF identification; in contrast, FRs are magnetically dominated, with a requirement of $|\sigma_r| < -0.5$ set here for their identification. The identification of AFs and SFRs is made here by imposing the above described criteria on the wavelet spectrograms of $\sigma_c$, $\sigma_r$, $\sigma_m$ in the frequency band 1.67--16.7 mHz (1 to 10 min) as visualised in Figs.~\ref{fig:example1} and \ref{fig:example2}, with the regions that meet all of the criteria outlined in cyan (AFs) and pink (SFRs) contours in the bottom panels. 

Figure~\ref{fig:shock_FRAC} shows how the 30- and 60-event running medians (grey and black lines, respectively) of the occurrence percentage of AFs and SFRs in the upstream and downstream depend on the shock characteristics. The occurrence percentages are calculated here as the number of bins in the wavelet spectrograms that meet the AF or SFR criteria above divided by the total number of bins. The shaded areas indicate the interquartile range for the 30-event medians.

Firstly, Fig.~\ref{fig:shock_FRAC} reveals that there are considerably more intervals fulfilling the AF criteria than the SFR criteria, and that the shock upstream has more AFs but fewer SFRs than the downstream (note the different y-axis scales in the figure). The mean occurrence percentages of AFs are 14.8\% and 10.8\%, and for SFRs 1.1\% and 1.7\% in the upstream and downstream, respectively (Table \ref{table:means}). The occurrence of AFs both in the upstream and downstream peak with $\upbeta_{\mathrm{u}} \sim 1-2$, and is lowest for the most perpendicular shocks. There is no obvious trend with the gas compression ratio, although the occurrence of AFs drops significantly at $r_\mathrm{g} \gtrsim 3$.  In the downstream, AF occurrence clearly increases with increasing $\Delta$V, while in the upstream it peaks for $\Delta$V $\sim 100 -150$ kms$^{-1}$ and for quasi-parallel shocks. The bottom two rows show that the SFR occurrence both in the upstream and downstream increases with increasing $\upbeta_{\mathrm{u}}$. In the upstream there is a weak declining trend in SFR occurrence with increasing shock velocity jump, while in the downstream the SFRs occurrence increases with increasing $r_\mathrm{g}$. In contrast to AFs, there is no trend with the shock angle. 

The heatmaps in Fig.~\ref{fig:change_FRAC} show the relative change in the AF and SFR percentage occurrences. Similar to Fig.~\ref{fig:change}, relative occurrence rates are calculated by dividing the number of events in a bin with the total number of events in the corresponding shock parameter range. The 40-event running medians and upper and lower quartiles are also shown. In agreement with Fig.~\ref{fig:shock_FRAC}, the individual events show that the change in the percentage occurrence for AFs from the upstream to downstream is on average negative, i.e. the upstream has more AFs than the downstream, while more SFRs occur in the downstream; the opposite trends are observed in some individual events, however. 

\begin{figure*}[ht]
\centering
\includegraphics[width=0.80\linewidth]{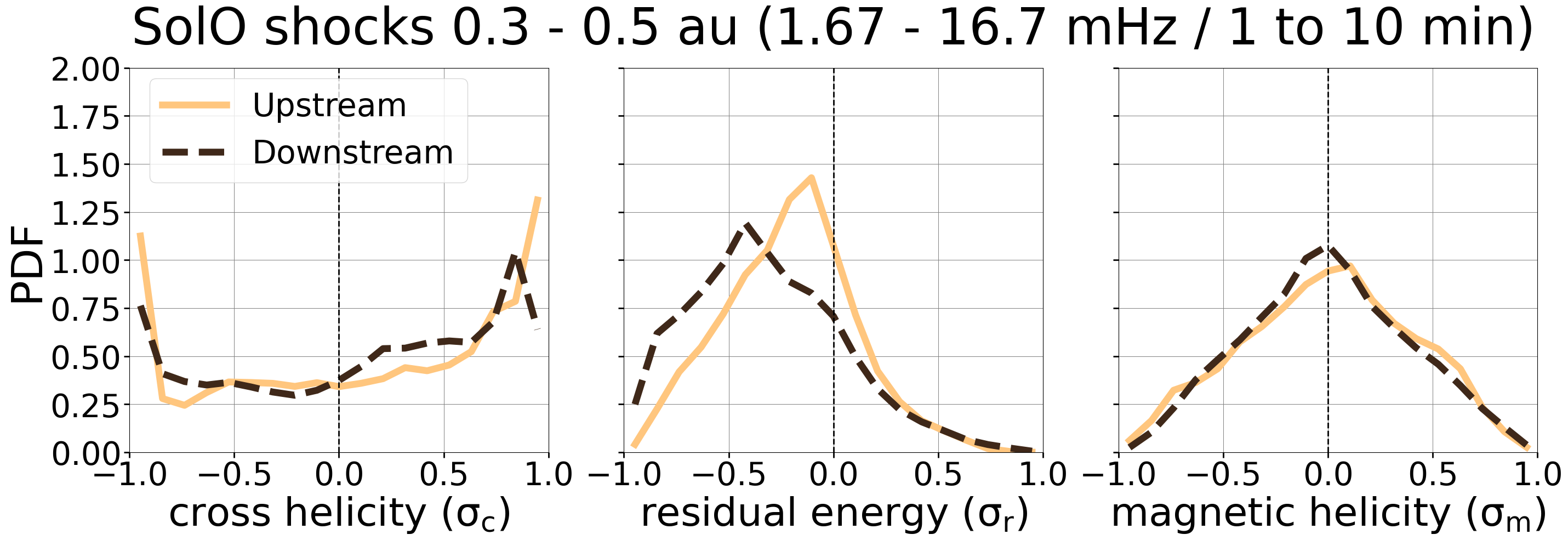}
\caption{PDFs of cross-helicity, residual energy and magnetic helicity for seven \textit{Solar Orbiter} shocks observed below 0.5 au.}
   \label{fig:SOLO_shocks}
\end{figure*}

\begin{figure}[ht]
\centering
\includegraphics[width=\linewidth]{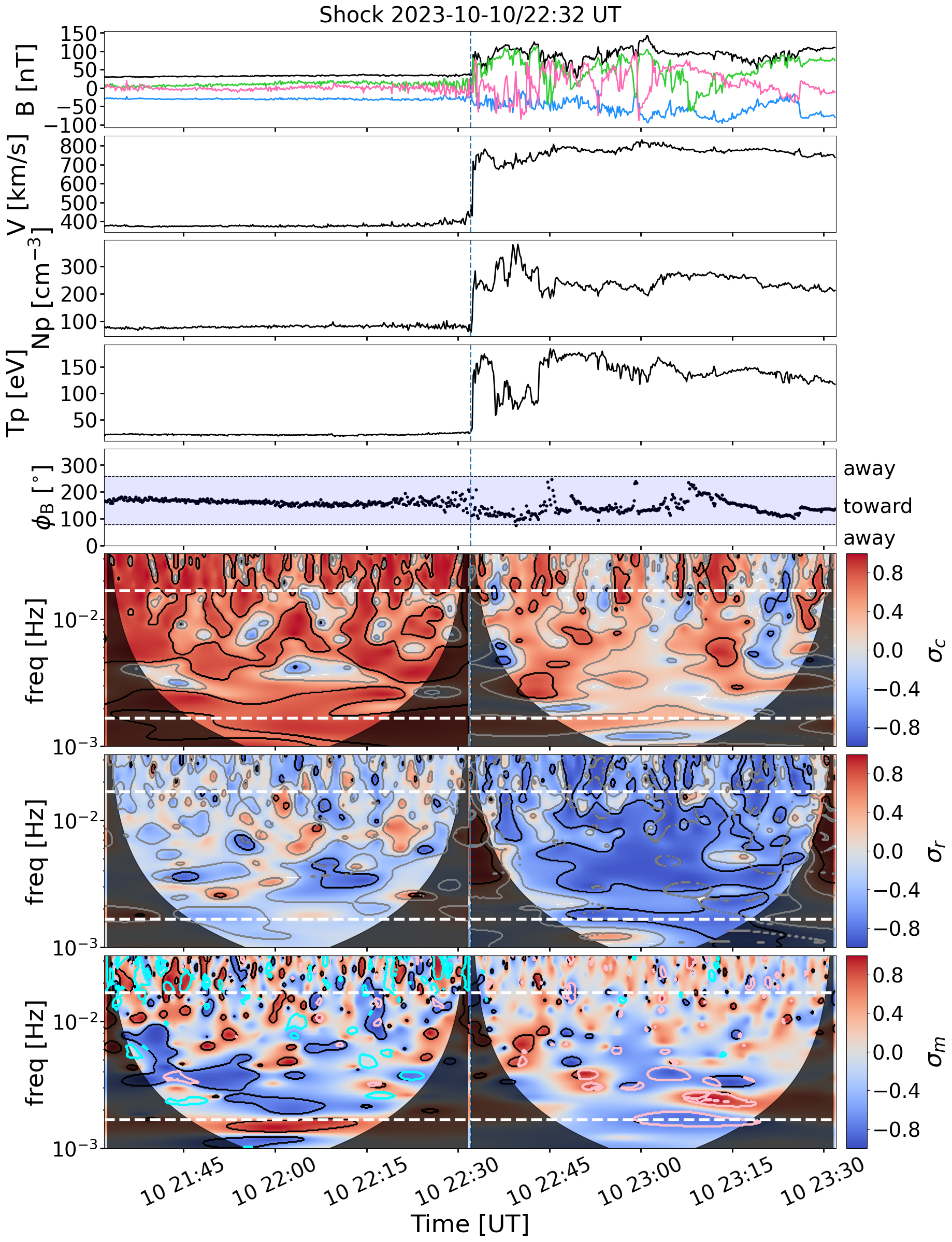}
\caption{A shock observed by \textit{Solar Orbiter} on 10 October 2023. The panel layout is the same as in Figs.~\ref{fig:example1} and \ref{fig:example2}.}
\label{fig:exampleSOLO}
\end{figure}

The tendency for the upstream to have more AFs than the downstream is greatest for shocks associated with $r_\mathrm{g} \gtrsim 2.5$, $\upbeta_{\mathrm{u}} \lesssim 2$ and $\Delta$V $\gtrsim 100$ kms$^{-1}$. SFRs are most abundant in the downstream compared to the upstream when the upstream beta and velocity jumps are large.  

The decrease in the AF occurrence and increase in the FR occurrence across the shock are obvious for our example event shown in Fig.~\ref{fig:example2}. In the upstream region, there are several regions that meet the criteria for AFs (surrounded by cyan contours) for all frequencies, while in the downstream AFs have largely disappeared. There are several regions meeting the criteria for SFRs (surrounded by pink contours). This example was a parallel shock with a large gas compression ratio. The example event in Fig.~\ref{fig:example1} is a case where the bins fulfilling the FR criteria diminish from upstream to downstream. This shock had a similar speed jump but was almost perpendicular.

\begin{table*}[!h] \caption{Shock parameters for the seven analysed shocks observed by \textit{Solar Orbiter}. The bottom row gives the mean values with standard deviations in parenthesis.}
\centering
\begin{tabular}{c c c c c c}
\hline\hline
 Date \& Time  & r     & $r_\mathrm{g}$ & $\upbeta_{\mathrm{u}}$ & $\Delta$V  & $\theta_\mathrm{Bn}$ \\ 
 $[$UT$]$     & [au]   &       &     & [$^{\circ}$] & [kms$^{-1}$] \\
\hline 
\hline
2022-03-08 14:46  & 0.48 & 2.32 & 0.32 & 59  & 53.1 \\
2022-03-11 19:52  & 0.44 & 2.87 & 0.29 & 281 & 25.6 \\
2022-04-03 04:52  & 0.36 & 2.33 & 0.74 & 141 & 42.2 \\
2023-04-10 04:33  & 0.29 & 1.83 & 1.26 & 125 & 23.4 \\
2023-09-19 02:23  & 0.47 & 2.31 & 4.11 & 44 & 82.1 \\
2023-09-20 00:47  &	0.46 & 2.40 & 1.71 & 78 & 54.0 \\
2023-10-10 22:32  & 0.30 & 3.17 & 1.14 & 376 & 43.2 \\
\hline
Mean (std)  & & $2.46 \pm 0.40$ & $1.36 \pm 1.22$ & $158 \pm 116$ & $46.3 \pm 18.4$ \\
\hline 
\hline
\end{tabular} \label{table:means_solo}
\end{table*}

\subsection{Near--Sun observations}
\label{sec:nearsun}

Finally, we investigate whether similar changes in turbulence parameters are observed close to the Sun. Here we analyse seven shocks observed by \textit{Solar Orbiter} below 0.5~au (see Sect.~\ref{sec:data_events}). The values of the shock parameters for each shock, including their means and standard deviations, are given in Table \ref{table:means_solo}. The event-to-event variations in shock parameters are substantial, as indicated by the relatively large standard deviations; compared to the \textit{Wind} shocks, the seven \textit{Solar Orbiter} shocks have on average a somewhat higher gas compression ratio and lower upstream plasma beta, while four are associated with substantially larger velocity jumps. The shocks are also more parallel, with four being in the quasi-parallel regime ($\theta_\mathrm{Bn} < 45^{\circ}$). 

The combined PDFs of $\sigma_c$, $\sigma_r$ and $\sigma_m$ for the seven shocks are shown in Fig.~\ref{fig:SOLO_shocks}. The results show an overall similar behaviour as at 1~au, namely, that the fluctuations become more balanced and magnetically dominated from upstream to downstream. The residual energy PDFs most resemble that of the \textit{Wind} upper quartile population, while the cross-helicity PDFs shows a less obvious decrease in balance between upstream and downstream. 

Figure~\ref{fig:exampleSOLO} shows an example event from \textit{Solar Orbiter}. The spacecraft detected a shock on 10 October 2023 22:32 UT when the spacecraft was 0.3~au from the Sun. The shock was just at the threshold between being quasi-parallel and quasi-perpendicular with $\theta_{Bn} = 43^{\circ}$, and the gas compression ratio had a relatively large value of 3.2. The speed jumped from about 390 kms$^{-1}$ to almost 800 kms$^{-1}$ at the shock. The \textit{Solar Orbiter} data is given in RTN (Radial - Tangential - Normal) coordinates. In the RTN system at 1~au, the outward (away) polarity is at clock angles $\phi$ below $45^{\circ}$ or over $225^{\circ}$, while in the towards sector $\phi$ is in the interval $45^{\circ} - 225^{\circ}$; for the \textit{Solar Orbiter} shocks, these angles have been adjusted relative to the local Parker-spiral winding angle given by the formula in Sect.~\ref{sec:calc_turb}, with the spiral angle becoming progressively smaller as the heliocentric distance reduces.

In the example event, the IMF is consistently in the towards sector. Thus the positive cross-helicity, which indicates waves travelling anti-parallel to the magnetic field, corresponds to the typically observed anti-sunward propagation. The upstream is imbalanced with average $|\sigma_c| = 0.60$. In the downstream, the there are some negative cross-helicity patches suggesting the presence of sunward propagating waves as well, and fluctuations are less imbalanced with average $|\sigma_c| = 0.36$. The residual energy clearly becomes more negative from upstream to downstream, with the average $\sigma_r$ changing from -0.18 to -0.62. The last panel shows many areas meeting the criteria for Alfv\'en waves in the upstream, with only a few regions meeting the criteria for flux ropes. Unlike at 1~au, the upstream $\sigma_m$ does not average to zero, having a value -0.17. In the downstream, the trend is reversed with considerably more regions meeting the flux rope criteria but average $\sigma_m$ is now closer to zero (-0.069).

\section{Discussion}
\label{sec:discussion}

Our investigation of 1-h averaged inertial range values of normalised cross-helicity, $\sigma_c$, residual energy, $\sigma_r$, and magnetic helicity, $\sigma_m$, in the vicinity of interplanetary shocks are in overall agreement with $\sim $1~au solar wind values reported in previous literature (see Sect.~\ref{sec:introduction}); we found that, on average, fluctuations are more frequently imbalanced ($| \langle \sigma_c \rangle | >  0.5$) than balanced ($| \langle \sigma_c \rangle| < 0.3$), outward fluctuation energy dominates (rectified cross-helicity $\sigma_c^* > 0$), the magnetic energy of fluctuations clearly exceeds the kinetic energy (negative $\sigma_r$), and magnetic helicity values average to zero. Cross-helicities exhibited a large variability, consistent with previous studies, which have also shown that their values vary with heliospheric distance and latitude, solar cycle phase, solar wind properties and large-scale structure \citep[e.g.,][]{Bruno2013,Soljento2023,Good2023,Perri2010,Amicis2007,Amicis2011,Chen2020,Bavassano1998}. As our study spans over 2.5 solar cycles, a significant spread in turbulence parameters could be at least partially explained by variations in the solar wind into which the shocks propagate. 

However, most interplanetary shocks observed in the ecliptic plane propagate into relatively slow solar wind \citep[e.g.][]{Kilpua2015}. This is the case also for our data set, which has a mean upstream solar wind speed of 397~kms$^{-1}$ and standard deviation of 84.7 kms$^{-1}$. Although anti-sunward imbalance in $\sigma_c$ is particularly strong in the fast solar wind \citep[e.g.,][]{Matthaeus1982,Bavassano1998}, it has been detected in the slow solar wind at 1~au \citep[e.g.,][]{Amicis2011,Amicis2021}. The outward imbalance likely results from anti-sunward propagating Alfv\'enic fluctuations. As previously mentioned, Alfv\'en waves (AFs) are locally characterised by high $|\sigma_c|$ and equipartition of $\sigma_r$, with an anti-sunward imbalance typically observed when averaging over longer intervals.

It is well known that shocks in space plasmas (e.g. planetary bow shocks, interplanetary shocks) generate AFs that propagate upstream via instabilities induced by shock-accelerated ions \citep{Lee1982}. In the case of forward propagating interplanetary shocks such as those analysed in the present work, an enhanced presence of anti-sunward AFs is thus expected to increase the imbalance of $\sigma_c$. Theoretical work indicates that quasi-parallel and strong shocks are most effective in self-generating AFs \citep{Vainio2005}. This is in agreement with our findings on how $\sigma_c$ and $\sigma_r$ depend on shock parameters, with further evidence of this phenomenon provided by our investigation of the presence of AFs using criteria imposed on $\sigma_c$, $\sigma_r$ and $\sigma_m$. The occurrence of intervals fulfilling the criteria for AFs was clearly higher both in the upstream and downstream of quasi-parallel shocks than quasi-perpendicular shocks. While imbalance in $\sigma_c$ and occurrence of AF-like intervals first increased with increasing gas compression ratio and shock velocity jump, at more extreme values the fluctuations again become more balanced, less equipartitioned and the occurrence of AF-intervals decreased, particularly in the downstream. This could be due to the strongest shocks generating compressive fluctuations that reduce Alfv\'enicity. The increased imbalance with increasing velocity jump could also result if the high velocity-jump shocks tend to have higher upstream solar wind speeds (and thus higher imbalance). However, the linear Pearson correlation coefficient calculated between the mean shock velocity jump and the upstream solar wind speed for our data set is only 0.33, indicating a weak correlation. 

Our study also confirms some previous findings related to how turbulence parameters change across the shock and gives new insight on how their changes depend on the shock parameters. In agreement with previous studies \citep[e.g.,][see Sect.~\ref{sec:introduction}]{Zhao2021,Borovsky2020,Good2022,Soljento2023} we found that fluctuations in the shock upstream tend to be more equipartitioned and imbalanced than in the downstream. The greater balance in the downstream occurred particularly for shocks with large velocity jumps and gas compression ratios. This could result from (i) a relatively more efficient transmission, generation or amplification of the sunward fluctuations, (ii) a relatively less efficient transmission, generation or amplification of the anti-sunward fluctuations, or (iii) a significant development of compressive fluctuations in the downstream. Residual energy becomes less equipartitioned downstream with higher power in magnetic field fluctuations, possibly suggesting that new, non-Alfv\'enic fluctuations are created. Our findings are also consistent with the results of \citet{Pitna2023P}, who find that the residual energy decreases from upstream to downstream when the amplified fluctuations are modelled as 2D turbulence.

However, we observed a substantial number of events in which fluctuations became more imbalanced and equipartitioned from upstream to downstream. These cases were mostly related to shocks with smaller gas compression ratios and velocity jumps, in agreement with the assumption that such shocks do not generate AFs effectively. One scenario in which imbalance could increase across the shock is where AFs generated by the shock in the upstream propagate more slowly than the shock, causing them to be left behind and subsequently overtaken by the shock, thus eventually becoming downstream waves.

The PDFs of residual energy and rectified cross-helicity in this study are in general agreement with those reported in \cite{Soljento2023}, who made comparisons between the upstream wind, sheath and ejecta associated with 70 shocks detected by the \textit{Wind} spacecraft. The authors found that $\sigma_c^*$ values were clearly more balanced in the sheath than in the preceding solar wind, while the differences in $\sigma_r$ PDFs between different regions were relatively small. This could be because they did not separate the events according to the shock parameters and considered the whole sheath instead of the 1-h intervals upstream and downstream.

Our identification of more small-scale flux ropes (SFRs) in the shock downstream and more AFs in the upstream is also consistent with a statistical study by \cite{Ruohotie2022} that covers CME-driven sheath regions, with their high frequency range (1-10 mHz) corresponding roughly to the range in our study. As discussed above, periods fulfilling the AF criteria were the most abundant in the upstream compared to downstream for shocks with large velocity jumps, gas compression ratios and quasi-parallel shock configurations, in agreement with theories of shock self-generation of AFs. The observed trend that the occurrence of AFs peaks strongly (both upstream and downstream) with $\upbeta_{\mathrm{u}} \sim 1-2$, and that the occurrence of SFRs in the downstream increases with increasing $\upbeta_{\mathrm{u}}$ are interesting. The trend for AFs could be explained by their damping in high-beta plasma \citep[e.g.,][]{Volk1982,Hollweg1971,Squire2017NJPh}. SFRs can result from magnetic reconnection and be self-generated from the turbulence cascade \citep[e.g.,][]{Zheng2018}. In addition, the passage of interplanetary shock past current sheets in the solar wind may trigger reconnection in them and breakdown to magnetic islands or flux ropes \citep[e.g.,][]{Odstrcil1997,Nakanotani2021}. This could explain the observed trend between $\upbeta_{\mathrm{u}}$ and the SFR occurrence, as well as why for high $\upbeta_{\mathrm{u}}$ shocks the occurrence of SFRs increases from upstream to downstream. In addition, higher intermittency and steeper spectral indices found in CME-driven sheaths compared to the upstream \citep{Kilpua2021Fr} are consistent with the idea that current sheets are generated in the shock downstream. Furthermore, reconnecting structures are frequently observed downstream of the Earth's bow shock both in simulations and observations, without strong dependence on shock orientation or Mach number \citep[e.g.,][]{Gingell2020,Gingell2023}. The seven \textit{Solar Orbiter} shocks detected below 0.5~au showed qualitatively similar behaviour as 1~au shocks, likewise featuring a decrease in AWs and increase of SFRs from upstream to downstream. 

\conclusions

We have performed a statistical analysis of the inertial-range normalised cross-helicity, residual energy and magnetic helicity upstream and downstream of 371 interplanetary shocks waves detected by the \textit{Wind} spacecraft at 1~au and seven shocks detected by \textit{Solar Orbiter} at less than 0.5~au. We found in the shock vicinity that average residual energies are negative (magnetic energy dominates), magnetic helicities average to zero and cross-helicities have a large spread with a preference for anti-sunward imbalance, in agreement with general solar wind observations. Our study shows that shock transitions may significantly affect the investigated turbulence parameters. While weak shocks do not typically alter turbulent properties in a significant way, shocks with large gas compression ratios and velocity jumps have a clear association with downstream fluctuations that are more balanced and less equipartitioned. Consistent with shock theories, the upstreams of quasi-parallel shocks were found to be the most Alfv\'enic (i.e. having the most imbalanced and equipartitioned fluctuations). Magnetic helicities were largely unaffected by the shock, averaging close to zero in all cases. The dependence on the upstream plasma beta of the occurrence of periods containing Alfv\'enic fluctuations and flux ropes suggests a relation to physical processes occurring at the shock, namely, the damping of Alfv\'enic fluctuations in high-beta plasma and triggering of reconnection in current sheets passing through the shock. The shock observations at heliocentric distances below 0.5~au showed qualitatively similar results as at 1 au. The linking of trends in turbulent properties found in this study to shock acceleration of energetic particles will be an interesting avenue of future research, given that energisation and acceleration of particles at collisionless shocks is an ubiquitous process.  

\dataavailability{The solar wind data used in this study are available from the NASA Goddard Space Flight Center Coordinated Data Analysis Web (CDAWeb; \url{http://cdaweb.gsfc.nasa.gov/}).} 

\authorcontribution{EK performed the data analysis and prepared the figures. All authors have contributed to the writing of the manuscript and interpretation of the results.}

\competinginterests{No competing interests are present.}

\begin{acknowledgements}
We acknowledge the Finnish Centre of Excellence in Research of Sustainable Space (Research Council of Finland grants 352850 and 352847). EK acknowledges the ERC under the European Union's Horizon 2020 Research and Innovation Programme Project SolMAG 724391.  We also acknowledge funding from the European Union’s Horizon 2020 research and innovation programme under grant agreement No. 101004159 (SERPENTINE). SG is supported by Research Council of Finland grants 338486, 346612 and 359914 (INERTUM). JS is supported by the Maili Autio Fund of the Finnish Cultural Foundation, grants 00231092 and 00242648.
\end{acknowledgements}

\bibliographystyle{copernicus}
\bibliography{reference}

\appendix

\section{Figures with full range of values}

\begin{figure*}[]
\includegraphics[width=0.99\linewidth]{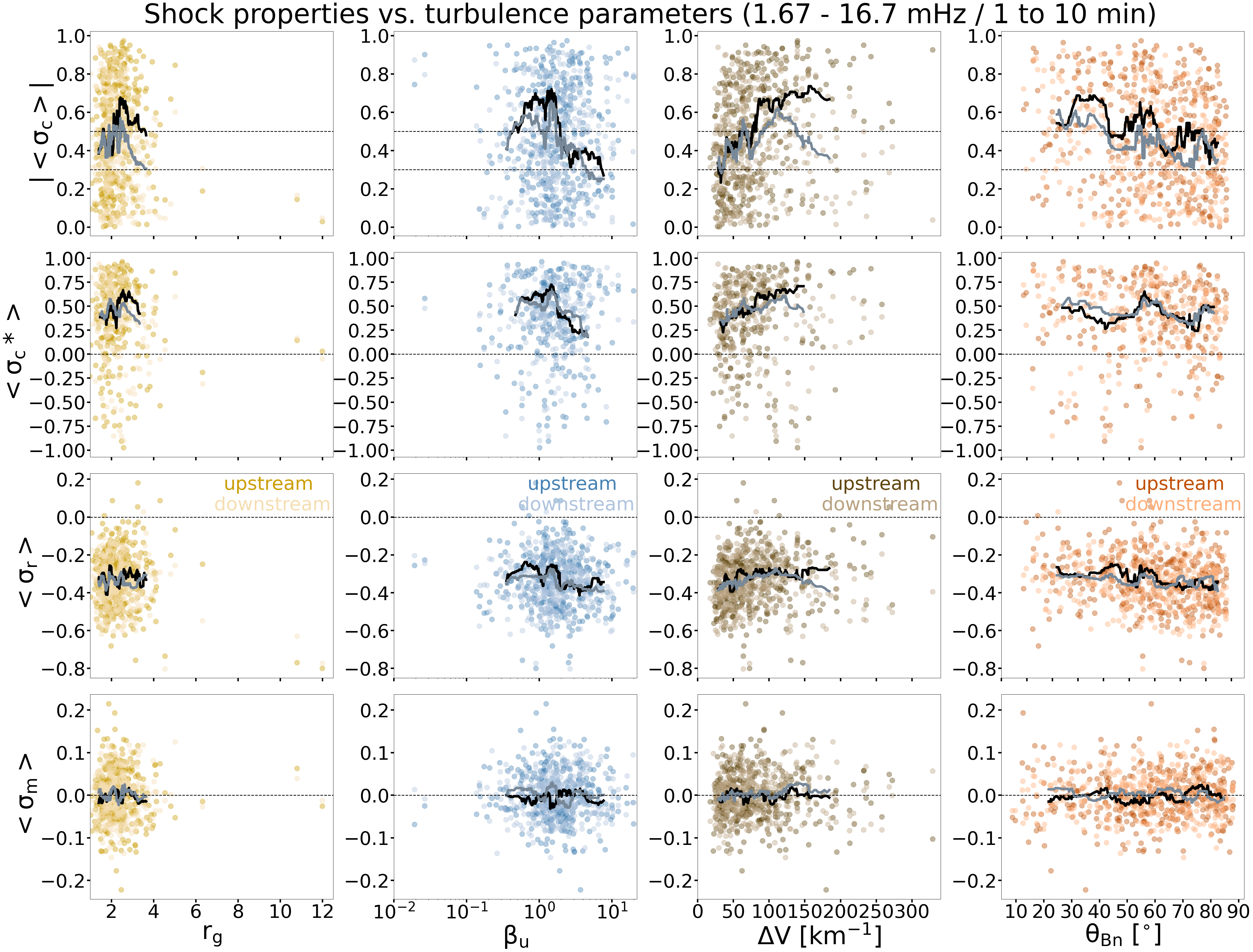}
\caption{Version of Fig.~\ref{fig:shock_turbulence} with the full x-axis ranges shown.}
   \label{fig:shock_turbulence_S1}
\end{figure*}

\begin{figure*}[]
\includegraphics[width=0.99\linewidth]{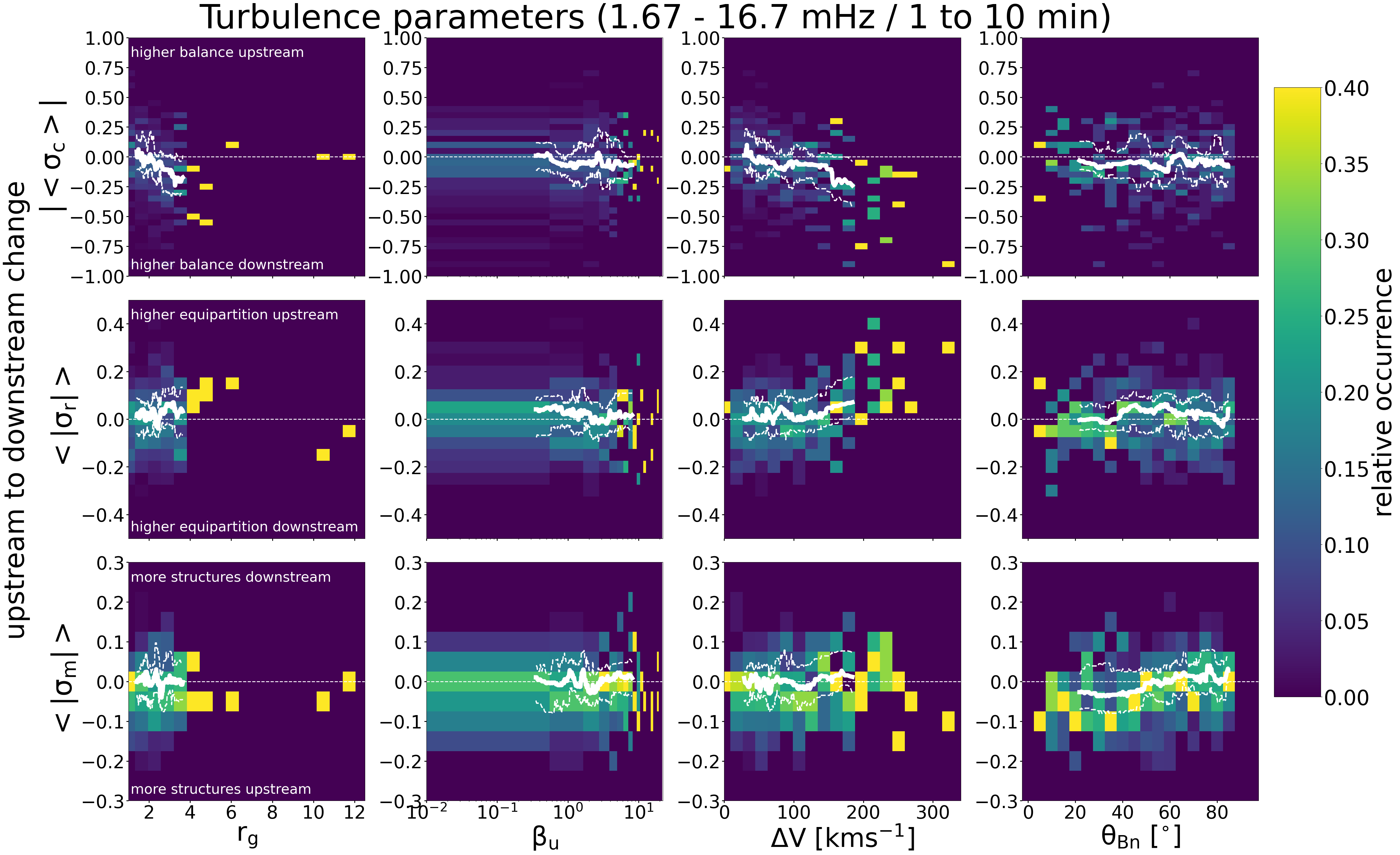}
\caption{Version of Fig.~\ref{fig:change} with the full x-axis ranges shown.}
   \label{fig:change_S2}
\end{figure*}

\begin{figure*}[]
\includegraphics[width=0.99\linewidth]{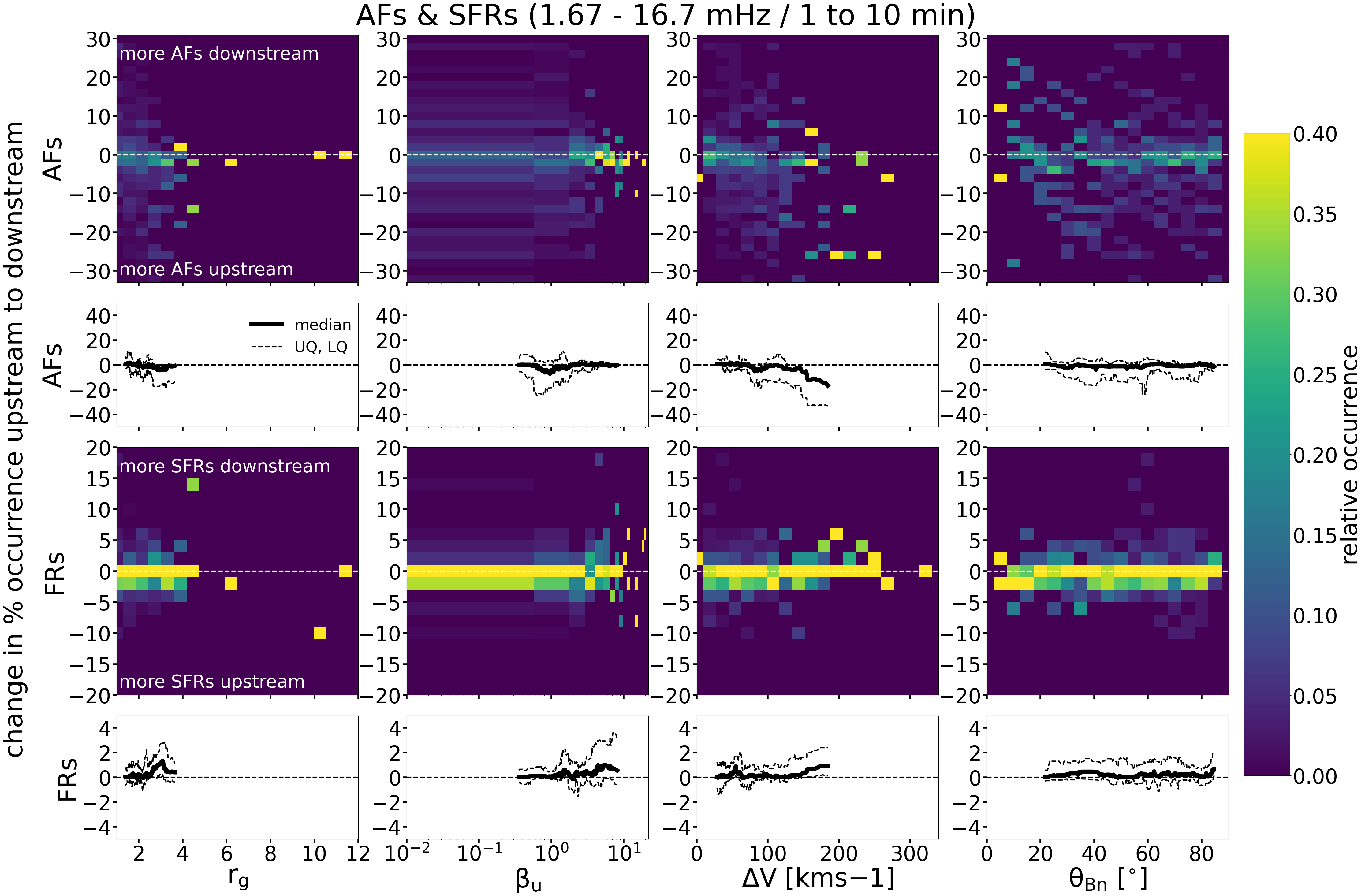}
\caption{Version of Fig.~\ref{fig:change_FRAC} with the full x-axis ranges shown.}
   \label{fig:change_S3}
\end{figure*}

\end{document}